\documentclass{aastex61}

\usepackage{longtable} 
\usepackage{graphicx}

\def \hi    	{H\,{\sc i}}
\def \civ    	{C\,{\sc iv}}
\def \oiv    	{O\,{\sc iv}}
\def \mgii    	{Mg\,{\sc ii}}
\def \siiv   	{Si\,{\sc iv}}
\def \siv    	{S\,{\sc iv}}
\def \caii   	{Ca\,{\sc ii}}
\def \caxvii   	{Ca\,{\sc xvii}}
\def \feix		{Fe\,{\sc ix}}
\def \fexvi		{Fe\,{\sc xvi}}
\def \fexviii  	{Fe\,{\sc xviii}}
\def \fexix  	{Fe\,{\sc xix}}
\def \aia	 	{{\it AIA}}
\def \chianti	{{\it Chianti}}
\def \iris   	{{\it IRIS}}
\def \arcsec 	{\hbox{$^{\prime\prime}$}}

\received{12/05/2018}
\revised{01/09/2019}
\accepted{01/12/2019}

\submitjournal{ApJ}

\shorttitle{Formation of Emission Lines Observed by IRIS}
\shortauthors{Bradshaw and Testa}
 
\begin{document}

\title{Quantifying the Influence of Key Physical Processes on the Formation of Emission Lines Observed by IRIS: I. Non-Equilibrium Ionization and Density-Dependent Rates}
 
\correspondingauthor{Stephen J. Bradshaw}
\email{stephen.bradshaw@rice.edu, ptesta@cfa.harvard.edu}

\author{Stephen J. Bradshaw}
\affiliation{Department of Physics and Astronomy, Rice University, Houston, TX 77005, USA}

\author{Paola Testa}
\affiliation{Smithsonian Astrophysical Observatory, 60 Garden Street, MS 58, Cambridge, MA 02138, USA}
 

\begin{abstract}
In the work described here we investigate atomic processes leading to the formation of emission lines within the \iris\ wavelength range at temperatures near $10^5$~K. We focus on (1) non-equilibrium and (2) density-dependent effects influencing the formation and radiative properties of \siv\ and \oiv. These two effects have significant impacts on spectroscopic diagnostic measurements of quantities associated with the plasma that emission lines from \siv\ and \oiv\ provide. We demonstrate this by examining nanoflare-based coronal heating to determine what the detectable signatures are in transition region emission. A detailed comparison between predictions from numerical experiments and several sets of observational data is presented to show how one can ascertain when non-equilibrium ionization and/or density-dependent atomic processes are important for diagnosing nanoflare properties, the magnitude of their contribution, and what information can be reliably extracted from the spectral data. Our key findings are the following. (1) The S/O intensity ratio is a powerful diagnostic of non-equilibrium ionization. (2) Non-equilibrium ionization has a strong effect on the observed line intensities even in the case of relatively weak nanoflare heating. (3) The density-dependence of atomic rate coefficients is only important when the ion population is out of equilibrium. (4) In the sample of active regions we examined, weak nanoflares coupled with non-equilibrium ionization and density-dependent atomic rates were required to explain the observed properties (e.g. the S/O intensity ratios). (5) Enhanced S/O intensity ratios cannot be due solely to the heating strength and must depend on other processes (e.g. heating frequency, non-Maxwellian distributions).
\end{abstract}

 

\section{Introduction}
\label{introduction}

The origin of the million degree solar corona, and the hot coronae of other magnetically active stars, is one of the central problems in modern astrophysics. Space-based observatories with instruments designed to observe the solar atmosphere with high spatial resolution and fast cadence, such as {\it Hinode} \citep{Kosugi2007}, the {\it Solar Dynamics Observatory} \citep[{\it SDO};][]{Pesnell2012}, the {\it High Resolution Coronal Imager} \citep[{\it HiC};][]{Cirtain2013}, and the {\it Interface Region Imaging Spectrograph} \citep[\iris;][]{DePontieu2014}, have revealed its strongly dynamic nature at small scales, implying an impulsive mechanism heating the plasma to high temperatures. The physical mechanisms that have been proposed to satisfy the requirements for coronal heating range from direct-deposits that involve the build-up and release of energy \citep[e.g.~magnetic reconnection;][]{Parker1988}, to more periodic processes \citep[e.g.~Alfv\'{e}n~wave~heating;][]{vanBallegooijen2011,AsgariTarghi2013}. The term {\it nanoflare} was first used by \cite{Parker1988} to describe coronal heating by magnetic reconnection on small spatial scales, but should now be considered an umbrella term for any impulsive heating event independent of the driving mechanism (e.g. reconnection and~/~or waves) and energy range \citep{Klimchuk2015}. {\it Nanoflare} will be used in this modern context here.

In the work reported here we have undertaken a quantitative study of nanoflare-based coronal heating and its detectable signatures in transition region emission. We have conducted numerical and forward modeling to predict and examine the properties of key spectral lines that are observed by \iris. Atomic processes that are expected to be of significant importance in the regions of line formation have been included in the modeling to determine which of those processes are most critical for diagnosing the nanoflare properties, and what information can be extracted from the spectral data. In this article we focus on the coupling of non-equilibrium ionization and density-dependent collisional ionization and recombination rates. A subsequent article will address coupled non-equilibrium ionization and non-thermal electron distributions. We have compared our numerical results with several sets of observational data and used our analysis to draw some conclusions concerning the nature of nanoflares in the observed regions. While we have focused on active regions for the present work, there is no particular reason to suppose that our findings and nanoflares, in one or more guise, are any less applicable to the quiet or diffuse corona.

In the remainder of this Introduction we discuss the difficulties associated with directly observing nanoflaring signatures in the corona (Section~\ref{introCorona}), possible indirect nanoflare signatures in the transition region that arise as a consequence of the directly coupled nature of these two layers of the solar atmosphere (Section~\ref{introTR}), and the key atomic processes governing the formation of emission lines relevant to our study (Section~\ref{introAtomic}). The rest of the article: describes our numerical and forward-modeling approaches (Section~\ref{model}) and our observational data sets (Section~\ref{observations}); provides an analysis of our results (Section~\ref{results}); and closes with a discussion of our findings and conclusions (Section~\ref{discon}).

\subsection{Signatures of Impulsive Heating in Coronal Emission}
\label{introCorona}

The chief obstacle to understanding the physical nature of the coronal heating mechanism lies in the difficulty of detecting direct observational signatures that can be unambiguously connected to its properties. One of the major predictions of nanoflare models is a hot ($\approx5-10$~MK) component to the emission measure distribution during the first few seconds of heating \citep{Cargill1994,Cargill2014,Barnes2016a,Barnes2016b}. The high initial temperatures are required to drive sufficiently strong coronal cooling by thermal conduction and a return enthalpy flux into the corona, such that the observed factor of 10 density excess during radiative cooling at $\approx$ 1~MK is produced \citep{Aschwanden2001,Spadaro2003,Winebarger2003}. The detection of plasma at temperatures in the range $5-10$~MK would be very strong evidence in favor of nanoflare heating because the demands on steady heating to maintain plasma at that temperature are energetically too extreme and its observational consequences would be easily detectable \citep{Schmelz2009a}. For example: given a fixed length $2L$, a coronal loop subject to steady heating at 5~MK would be 25 times more dense, 625 times brighter, and require a volumetric heating rate 280 times greater than a coronal loop heated steadily at 1~MK \citep[based on an estimate using the RTV scaling laws;][]{Rosner1978}. A 10~MK coronal loop would be 100 times more dense, 10,000 times brighter, and require a volumetric heating rate 3160 times greater than the 1~MK case. As observing instruments become more sensitive and methods of data analysis more discriminating, evidence for hot plasma and nanoflare heating is beginning to emerge.

\cite{Schmelz2009a} and \cite{Reale2009a} analyzed {\it Hinode/XRT} \citep{Golub2007} observations of non-flaring solar active regions. Using combinations of filters sensitive to different energy ranges they found emission measures with maxima in the range $2-4$~MK, and components above 10~MK but almost three orders of magnitude weaker. This hot and faint component is consistent with the parameter space of impulsive heating properties, but inconsistent with steady heating. \cite{Reale2009b} compared observations of the non-flaring corona made with {\it Hinode/XRT} and {\it RHESSI} \citep{Lin2002} when the solar disk was dominated by a single active region. They derived emission measures from multi-filter {\it XRT} data, assessed their consistency with emission measures calculated from the {\it RHESSI} fluxes at a range of energies, and found a weak component of the emisison measure at $6-8$~MK. \citep[see also][]{McTiernan2009}. \citep{Schmelz2009b} extended their earlier study to also include {\it RHESSI} fluxes to better constrain their {\it XRT} emission measures, which resulted in shifting the hot component identified at 30~MK much closer to 10~MK.

\cite{Testa2011} analyzed {\it XRT} and {\it Hinode/EIS} \citep{Culhane2007} observations of an active region for which the emission measure peaked at 2~MK, and then selected three sub-regions; one cool and two hotter. In the hotter regions they found components of the emission measure that were about two orders of magnitude below the peak. \cite{Patsourakos2009} also used {\it EIS} data to calculate active region emission measures in the range $1-5$~MK, finding distributions that were flat or slowly increased to 3~MK before rapidly falling off towards higher temperatures, and demonstrated that the observed emission measures could be reproduced by impulsive heating models. \cite{Miceli2012} took data provided by the {\it SphinX} X-ray spectrometer \citep{Sylwester2008,Gburek2011} sensitive in the $1-15$~keV range. They found the observed spectrum could be reproduced by a two-component emission measure: a large, warm thermal component at 2.7~MK; and a minor, hot thermal component a 7~MK. The hot component was found to be $<0.1$~\% of the warm one. Notably, these data were taken close to solar minimum and the observed active regions were not particularly intense \citep[in common with the relatively cool active regions analyzed by][]{Testa2011}.

\cite{Testa2012} used {\it SDO/AIA} observations \citep{Lemen2012} and {\it EIS} spectral data to assess the ability of \aia\ (a narrowband instrument) to identify hot plasma at $6-8$~MK \citep[the equilibrium formation temperature of \fexviii; see also][]{Petralia2014}. They analyzed the emission from two active regions observed simultaneously by \aia\ and {\it EIS}, and compared 3-color images constructed by combining the 171~\AA, 335~\AA, and 94~\AA~channels with \caxvii\ (192.8~\AA) {\it EIS} raster images. The selected active regions appeared bright in the \caxvii\ emission in limited areas showing similarities with bright features in the \aia\ 94~\AA~channel, which is sensitive to a high temperature \fexviii\ emission line in addition to lower temperature emission. The distribution of emission in the \aia\ 171~\AA~channel was markedly different to the \caxvii\ emission observed by {\it EIS}, allowing the authors to confirm that the bright emission observed in the 94~\AA~channel coincident with the distribution of \caxvii\ emission was due to \fexviii. A combination of \aia\ channels can be used to confirm the presence of emission in hot channels, while using the cooler channels to discount the possibility of low temperature emission polluting the hot channels.

As evidence for hot sources in the non-flaring corona began to mount, rocket-borne instruments on sub-orbital trajectories made increasingly important contributions to the observational record. The {\it Extreme Ultra-violet Normal Incidence Spectrograph} ({\it EUNIS-13}) sounding rocket instrument observed faint, but pervasive, \fexix\ 592.2~\AA~line emission indicating the presence of plasma heated to temperates of $\approx$~9~MK \citep{Brosius2014}. \cite{Caspi2015} measured SXR irradiance (0.5 to 5~keV) using the Amptek~X123-SDD spectrometer flown on two sounding rockets. They observed SXR emission orders of magnitude greater than during solar minimum when weak activity was present and spectra consistent with $5-10$~MK plasma, described by power-laws with indices of $\approx6$ predicted by nanoflare heating models. The {\it Focusing Optics X-ray Solar Imager} ({\it FOXSI}), flown on three sub-orbital rocket flights, is an instrument designed to directly focus hard X-rays (HXRs) rather than indirectly image their sources, which greatly increases the sensitivity and dynamic range of detection \citep{Glesener2016}. \cite{Ishikawa2014} used HXR data provided by the second flight of the {\it FOXSI} instrument to constrain a differential emission measure (DEM) produced by {\it Hinode/EIS} and {\it XRT}. The {\it FOXSI}-constrained DEM discounted a significant emission component above 10~MK that appeared in the {\it Hinode}-only analysis, but was consistent with a weak, high-temperature component between the peak DEM, at $\approx$~2.24~MK, and 8~MK where the DEM fell four orders of magnitude below the peak.

The {\it Nuclear Spectroscopic Telescope Array} ({\it NuSTAR}) \citep{Harrison2013} is an astrophysics mission intended to explore active galactic nuclei activity, compact objects, supernovae, and blazars, at HXR energies in the range $3-79$~keV. In common with {\it FOXSI}, it utilizes direct focusing optics to capture sources, greatly enhancing its imaging capabilities. Though not its primary mission, every so often {\it NuSTAR} points at the Sun to provide a unique complement to the fleet of solar observatories \citep{Grefenstette2016}. \cite{Hannah2016} presented {\it NuSTAR} observations of quiescent active regions in the declining phase of the solar cycle, finding emission measures with an isothermal component in the range $3.1-4.4$~MK and placing constraints on a hot component between $5-12$~MK up to three orders of magnitude weaker.

The body of observational evidence for hot plasma in quiescent, non-flaring active regions collected to-date consistently points to the plasma being of relatively low-density, due to the weak intensities associated with it. The faintness of high temperature emission is in fact a prediction of nanoflare heating models. Prior to heating, coronal plasma may be expected to be in a cool and tenuous state, such that when it is impulsively heated the large amount of energy per particle forces it into a hot and tenuous state, far from thermal equilibrium. Thermal conduction is extremely efficient under these conditions, with the timescale for energy transport varying as $\tau_C \sim n_e T_e^{-5/2}$, and rapidly cools the plasma from its peak temperature before sufficient counts can be detected. The return enthalpy flux from the lower atmosphere fills the corona with denser, but cooler plasma. Thermal conduction is particularly effective at smoothing field-aligned gradients such that if sufficient counts to measure a direct heating signature were gathered, the resulting time-integration would destroy crucial information such as the spatial scale of the energy deposition \citep{Winebarger2004}. This is exacerbated if heating occurs on sub-resolution strands, or on multiple strands along the line-of-sight, because the observed emission is due to a superposition of strands in various stages of heating and cooling. The ionization timescales of the coronal ions that are assumed to characterize the plasma temperature can be much longer than the heating timescale, such that the ionization state lags the temperature change, forcing the ion populations out of equilibrium with the local temperature. The emission from these ``hot'' ions is then substantially weakened, or possibly non-existent, and thermal conduction cools the plasma before significant populations can accumulate \citep{Bradshaw2006,Reale2008,Bradshaw2009}. The {\it Marshall Grazing Incidence X-ray Spectrometer}~({\it MaGIXS}) \citep[][]{Kobayashi2011,Champey2016} rocket-borne instrument (launch date: summer 2019) has been designed to search for this predicted faint emission at X-ray wavelengths in the range $6-24$~\AA, with 5~arcsec spatial resolution to distinguish between different structures in quiescent, non-flaring active regions.

\subsection{Signatures of Impulsive Heating in Transition Region Emission}
\label{introTR}

Since direct signatures of coronal nanoflares are expected to be extremely difficult to detect then we may look to indirect signatures to provide evidence. \cite{Testa2014} measured variability in the intensity and velocity of the \iris\ \siiv\ (1402.77~\AA) emission line on timescales of $20-50$~seconds at the foot-points of hot and dynamic coronal loops. They used numerical modeling to infer that the observed activity was consistent with coronal nanoflares depositing energy on timescales of $10-30$~seconds, associated with beams of electrons travelling to the foot-points and depositing their energy below the layer of the atmosphere at which \siiv\ forms in equilibrium, driving upflows of $\approx15$~km/s \citep[see also][]{Polito2018}. \cite{Testa2013} observed strong variability on timescales as short as 15~seconds in active region moss with {\it HiC}. The moss is a manifestation of emission from the transition region foot-points of overlying hot loops and this observations presents a significant challenge to previous conclusions that the moss (and the hot loops by association) is heated in a mostly steady manner \citep{Antiochos2003,Brooks2009,Tripathi2010}. In the observations presented by \cite{Testa2013} the configuration of the overlying loops evolved in a manner characteristic of slipping reconnection, suggesting an energy release located in the corona. The transition region and the corona are very strongly coupled. Information is exchanged between them via sound waves \citep{Cargill2013} which travel at high speeds in the hot corona and so the transition region can respond quickly to changing coronal conditions. In particular, the transition region is very sensitive to changes in the coronal pressure (e.g. due to heating) and so the variability of the moss emission is most likely connected to the timescale of energy release in the corona.

\iris\ observes the transition region at high spatial and temporal resolution (similar to {\it HiC}) and can detect small scale variability at loop foot-points, providing quantitative information about coronal heating. Before proceeding, we must be careful to define what is meant by the transition region in this context. It has been shown that the transition region is composed of two main components: the foot-points of overlying hot loops \citep[e.g.][]{Testa2016}; and short, low-lying loops that are highly dynamic on short timescales ($<1$~minute), which were previously observed as "unresolved fine structure" \citep[UFS:][]{Feldman1983,Feldman1999} and which \iris\ has now resolved \citep{Hansteen2014}. The latter structures seem to contribute most of the transition region emission, but do not extend to coronal altitudes or reach coronal temperatures. Hence, it is the former group of structures that we focus on here but one must be cautious when analyzing this emission. The dynamic nature of the plasma, the steep local gradients, and the overlying hot corona all combine to produce conditions where assumptions such as thermal equilibrium of the electron and ion populations may not be valid, which can strongly de-couple the resulting emission spectrum from the local temperature. In this article we focus on two atomic processes that may play a key role in determining the plasma conditions where emission lines detected by \iris\ are formed: non-equilibrium ionization and density dependence of ionization and recombination.

\subsection{Key Atomic Processes Encoding Signatures of Impulsive Heating}
\label{introAtomic}

Given the short timescales of variability observed in the transition region emission, and the steep temperature gradients and upflows of order $10$~km/s that are present; it is plausible that the ionization state experiences temperature changes on timescales faster than it is able to collisionally adjust, despite the relatively high densities in this layer of the solar atmosphere. For example, in a numerical model of coronal heating \cite{Olluri2013a,Olluri2013b} found significant non-equilibrium populations of \civ\ and \oiv, the latter contributing several important lines (at 1399.78~\AA, 1401.16~\AA, and 1404.81~\AA) to the wavelength sensitivity range of \iris. However, \cite{Hansteen1993} found \oiv\ populations close to equilibrium in a coronal nanoflare model, but did find strongly non-equilibrium populations of \civ. Collisional ionization and recombination rates are larger for \oiv\ than \civ, and the heating used in the latter study was relatively weak so that a coronal part of the loop locally exceeded 1~MK only briefly. Similarly, \cite{Doyle2013} found the ratio of \siiv\ (1394~\AA) to \oiv\ (1401~\AA) emission can change by up to a factor of 4 due to non-equilibrium ionization of \siiv\ alone but, again, the driving was relatively weak compared with the kind of pressure changes required in the corona to be consistent with the presence of hot ($\approx10$~MK) plasma.

Most commonly, the ionization and recombination rates that are provided for use in data analysis and modeling studies have been calculated in the low- (or zero) density limit \citep[e.g. the \chianti\ atomic database:][and references therein]{DelZanna2015,Dere1997}. The implicit assumption is that the electron density affects only the rate at which collisions occur, but has no further influence via, for example, the population of metastable energy levels or highly excited autoionizing states, which provide additional pathways to ionization as the density increases. This assumption may be tolerable in the relatively tenuous corona, but as one ventures deeper into the atmosphere its validity must be questioned. In the solar atmosphere, ionization is by collisional (electron-impact ionization) and recombination is by radiative and dielectronic recombination. Dielectronic recombination is the most important of these electron capture processes \citep{Burgess1964} for elements heavier than helium \citep{Nikolic2013}. Dielectronic recombination involves the following processes:

\begin{equation}
X^{z+1} (i) + e^- \leftrightarrow X^z (j,nl);
\label{eqn1}
\end{equation}

\begin{equation}
X^z (j,nl) \rightarrow X^{z+1} (k) + e^-;
\label{eqn2}
\end{equation}

\begin{equation}
X^z (j,nl) \rightarrow X^z (m,nl) + h \nu.
\label{eqn3}
\end{equation}

In the first step, represented by Equation~\ref{eqn1}, a $(z+1)$-times ionized atom of element $X$, in state $i$, captures an electron to become a $z$-times ionized atom in a doubly excited state represented by $j,nl$. The free electron is captured into orbital $nl$ and an inner (core) electron is excited by $i \rightarrow j$. At this point the ion has three choices: (1) if the inner electron radiatively decays directly to its initial state $j \rightarrow i$ then it autoionizes and releases the captured electron with its original energy (Equation~\ref{eqn1}); (2) if the inner electron decays to another excited state $j \rightarrow k$ then it autoionizes and releases the captured electron with less than its original energy (Equation~\ref{eqn2}); or (3) the ion can radiatively stabilize if the inner electron decays first by $j \rightarrow m$, emitting a photon of energy $h \nu_{j,m} < I_{nl}$ (where $I_{nl}$ is the ionization energy from level $nl$; Equation~\ref{eqn3}), and then to its final state via a radiative cascade.

In the limit of zero density, when the ion has radiatively stabilized, the outer electron in orbital $nl$ can only decay by a radiative cascade until it too reaches its final state. At this point, the dielectronic recombination process is complete. However, at finite electron densities there is the possibility for re-ionization, after the ion has radiatively stabilized, by electron impact. This can either happen directly or in a step-wise manner (before or during the radiative cascade). The consequence of this additional path to ionization is that the dielectronic recombination rate is {\it suppressed} relative to the zero density value at the same temperature, because it prevents the process from fully completing. This quenching of dielectronic recombination at high densities has the effect of shifting the ion population to lower temperatures in a gravitationally stratified atmosphere, with a positive temperature gradient (like the upper chromosphere~/~transition region~/~corona), than would be predicted by a calculation based on zero density rates \citep{Bradshaw2003,Polito2016}.

The quenching of dielectronic recombination has been calculated in the Collisional-Radiative modeling framework by \cite{Summers1974}. \cite{Nikolic2013} used the Collisional-Radiative results to estimate corrections for the density-dependent quenching that could be applied to the zero-density dielectronic recombination rates calculated by \cite{Badnell2003}. The {\it Atomic Data and Analysis Structure} \citep[{\it ADAS};][]{McWhirter1984,Summers2006} code adopts a Generalized Collisional-Radiative (GCR) model, which also accounts for the population of metastable states when calculating ionization and recombination between ion charge states.

\section{Numerical Model}
\label{model}

\subsection{Field-Aligned Hydrodynamics}
\label{modelHydro}

The numerical modeling component of this research was carried out using the {\it HYDrodynamics and RADiation} \citep[{\it HYDRAD};][]{Bradshaw2013} code and a sophisticated forward-modeling package \citep{Bradshaw2011,Bradshaw2016} to predict the signatures detectable in transition region emission when nanoflares energize the corona. We first describe the relevant details of {\it HYDRAD} that make it suitable for this study and then the particulars of the numerical experiments performed.

The low plasma $\beta$ in the solar atmosphere and the inhibition of cross-field mass and energy transport mean that we can treat each magnetic field line as an isolated atmosphere, solving for the plasma structure and evolution in the field-aligned direction as it responds to each nanoflare. {\it HYDRAD} solves the time-dependent equations for the evolution of mass, momentum, and energy for multi-fluid plasma (electrons, ions and neutrals) in arbitrary magnetic geometries taking account of the field-aligned gravitational acceleration and flux tube expansion (depending on e.g. magnetic field strength), and includes bulk transport (with shock capturing), thermal conduction (with heat flux limiting due to saturation and optional turbulence driven by magnetic fluctuations), viscous interactions, gravitational energy, Coulomb collisions, heating, and optically-thick radiation in the lower atmosphere transitioning to optically-thin radiation (lines and continuum) in the overlying atmosphere.

Optical depth effects in the lower solar atmosphere are treated by adopting the VAL~model~C chromosphere \citep{Vernazza1981} and a prescription for optically-thick radiative processes developed by \citet{Carlsson2012}. These authors have used calculations of time-dependent radiative transfer and snapshots from 2D~MHD simulations to find empirical formulae for the radiative cooling and heating of the chromosphere by \hi\, \mgii\ and \caii. In regions where the plasma is partially ionized the ionization and recombination of hydrogen, and the contribution of \hi\ to energy transport by thermal conduction \citep{Orrall1961}, are included in the energy equations solved by {\it HYDRAD}. This implementation of the lower atmosphere in {\it HYDRAD} was first used by \citet{Reep2013} and later by \citet{Reep2015,Reep2016}.

During nanoflare heating the plasma temperature may increase on timescales shorter than the ionization timescale. The collisionally-dominated ionization state cannot keep pace with the changing temperature and the result is a charge state characteristic of much lower temperatures \citep{Bradshaw2003,Bradshaw2006,Reale2008}. Similarly, strong flows that arise during heating can transport ions across steep temperature gradients before they can adjust to the changing conditions, decoupling the ionization state from the local temperature \citep{Raymond1978,Joslyn1979a,Joslyn1979b}. To properly capture this behavior, {\it HYDRAD} solves the time-dependent equations for the evolution of the ionization state of the radiating elements and can also fold the solutions back into the radiation calculation. The rate data for these calculations are provided by {\it ADAS}. The inclusion of this important physics for all of the most abundant elements in the solar atmosphere is one of the unique characteristics of the {\it HYDRAD} code.

The nanoflare energy input to the corona is treated via the addition of a source term to the electron energy equation that enables different properties of the heating and their consequences to be investigated. We have assumed that the heating mechanism energizes the electrons, which then exchange energy with the ions as these species thermally equilibrate via Coulomb collisions. Note that this assumption is implicit in single-fluid models, where heat conduction by electrons always dominates the thermal transport of energy in the plasma because $T_e = T_i$ everywhere.

In the present work we deposit energy in a spatially uniform manner into a coronal loop of total length (foot-point to foot-point) $2L=100$~Mm, which includes a chromosphere of height 2.2~Mm at each end of the loop (per the VAL~model~C). The key assumption underpinning the choice of uniform heating is that efficient (field-aligned) thermal conduction rapidly redistributes energy deposited in the corona through the atmosphere, such that time-integrated observables for uniform versus more localized heating are expected to be similar. The temporal envelope of each individual nanoflare is triangular and symmetric about the time of peak heating: the magnitude of the volumetric heating rate increases linearly to a maximum 60~s after onset and decays linearly to the background value 60~s later. To construct a nanoflare train we draw the maximum volumetric heating rate for each nanoflare randomly from a distribution described by a power-law with an index of -2.5. The period between nanoflares is determined by the energy of the next event; the physical reasoning underlying this choice is that the heating mechanism must re-energize (do work on) the magnetic field in order to build up sufficient stress~/~free energy and this process takes longer for more energetic nanoflares \citep{Cargill2014,Barnes2016b}.

\begin{figure}
	\plotone{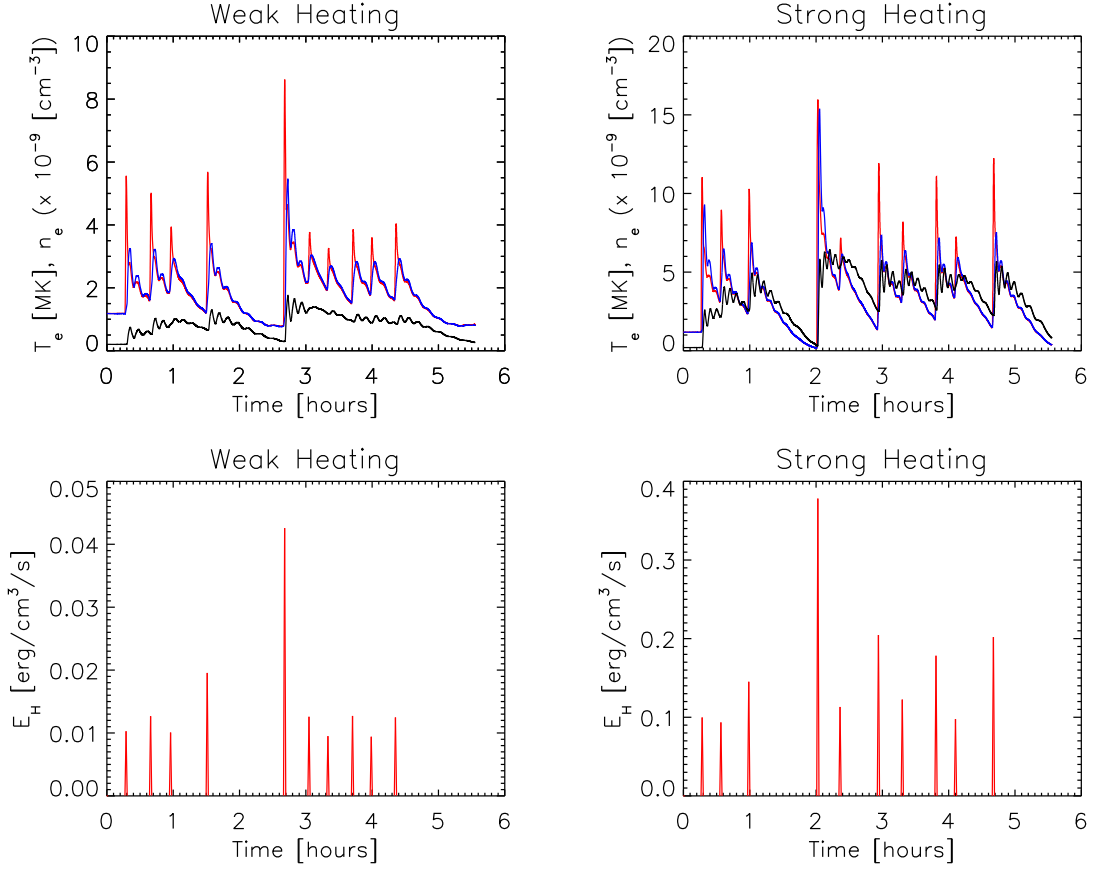}
	\caption{The temporal distribution of the nanoflares and the coronal response in the weak- and strong-heating scenarios. In the top two panels the red curve shows the electron temperature, the blue curve shows the ion temperature, and the black curve shows the electron number density.}
	\label{fig1}
\end{figure}

We have chosen two canonical nanoflare train-based numerical experiments to facilitate our investigation: a {\it weak}-heating and a {\it strong}-heating scenario. The average delay (waiting time) before the onset of nanoflares of the same relative energy (e.g. weakest, average, strongest) in each scenario is the same so that each experiment could be run for the same duration ($\approx 5$~hours of solar time). Figure~\ref{fig1} shows the temporal distribution of the nanoflares and the response of the coronal plasma (averaged over the upper 10\% of the corona) in the weak-heating scenario (left-hand column) and the strong-heating scenario (right-hand column).

\startlongtable
\begin{deluxetable}{c|ccc|ccc}
\tablecaption{The onset time, volumetric heating rate, and energy released per unit cross-section of the coronal loop for each nanoflare comprising the weak- and strong-heating trains.}
\tablehead{
 & & \colhead{Weak Heating} & & & \colhead{Strong Heating} & \\
\colhead{Nanoflare} & \colhead{Onset Time} & \colhead{Heating Rate} & \colhead{Energy Release} & \colhead{Onset Time} & \colhead{Heating Rate} & \colhead{Energy Release} \\
 & \colhead{[seconds]} & \colhead{[erg/cm$^3$/s]} & \colhead{[erg/cm$^2$]} & \colhead{[seconds]} & \colhead{[erg/cm$^3$/s]} & \colhead{[erg/cm$^2$]}
}
\startdata
1 & 981.68 & 0.0102 & $6.12\times10^{9}$ & 956.86 & 0.0997& $5.98\times10^{10}$ \\
2 & 2313.36& 0.0126 & $7.56\times10^{9}$ & 1973.63& 0.0934& $5.60\times10^{10}$ \\
3 & 3395.60& 0.0100 & $6.00\times10^{9}$ & 3486.66& 0.1451& $8.71\times10^{10}$ \\
4 & 5387.28& 0.0195 & $1.17\times10^{10}$& 7236.39& 0.3781& $2.27\times10^{11}$ \\
5 & 9588.96& 0.0425 & $2.55\times10^{10}$& 8441.78& 0.1131& $6.79\times10^{10}$ \\
6 & 10911.7& 0.0125 & $7.50\times10^{9}$ & 10525.3& 0.2045& $1.23\times10^{11}$ \\
7 & 11938.8& 0.0094 & $5.64\times10^{9}$ & 11820.9& 0.1225& $7.35\times10^{10}$ \\
8 & 13273.9& 0.0127 & $7.62\times10^{9}$ & 13651.7& 0.1782& $1.07\times10^{11}$ \\
9 & 14291.9& 0.0094 & $5.64\times10^{9}$ & 14708.5& 0.0976& $5.86\times10^{10}$ \\
10& 15604.6& 0.0124 & $7.44\times10^{9}$ & 16766.6& 0.2019& $1.21\times10^{11}$ \\
Total & & & $9.07\times10^{10}$ & & & $9.81\times10^{11}$ \\
\enddata
\label{table3}
\end{deluxetable}

The strong-heating case deposits approximately one order of magnitude more energy into the coronal loop than the weak-heating case, leading to an average temperature that is $\approx$ a factor of two greater and an average density $\approx$ four times greater. Assuming the heating mechanism re-energizes plasma on thin, sub-resolution magnetic strands of diameter on the order of 100~km ($10^7$~cm) then the total energy released per nanoflare in the weak-heating case is of order $10^{23}$ - $10^{24}$~ergs and of order $10^{24}$ - $10^{25}$~ergs in the strong-heating case.

A crucial aspect of solar atmosphere modeling is resolving the steep transition region temperature gradients so that the dynamic response of the atmosphere to heating is captured correctly \citep{Bradshaw2013}. This becomes more important with increasing loop temperature and so it is of particular relevance to the hot loops formed by the strong nanoflare train. Resolving the transition region is also absolutely critical to accurately forward modeling its emission, as will be discussed in Section~\ref{modelForward}. {\it HYDRAD} features adaptive mesh refinement that is capable of very fine resolution (to meter scales on modern hardware) and makes the code very efficient by ensuring that the spatial density of grid cells increases wherever needed, while keeping the total number of cells manageable. {\it HYDRAD} is a robust and well-tested code that has been used for a very large number of investigations in the published literature.

\subsection{Forward Modeling}
\label{modelForward}

A key aspect of this investigation was to predict the variability of the transition region emission, in response to the energy deposition by the nanoflares trains, that could be observed by \iris. Of particular interest were the relatively low-temperature spectral lines formed towards the base of the transition region ($T_e\sim10^5$~K) emitted by \oiv\ (1399.78, 1401.16, and 1404.81~\AA), \siiv\ (1402.77~\AA), and \siv\ (1404.85 and 1406.06~\AA). The forward modeling necessary for this task was performed using a powerful spectral synthesis code \citep{Bradshaw2011,Bradshaw2016} which incorporates tabulated atomic data from the \chianti\ database, and a set of wavelength-resolved response functions for a range of imaging and spectroscopic instruments provided by {\it SolarSoft}. {\it HYDRAD} outputs the species temperatures, pressures and number densities, bulk flow velocity, and ion populations as they evolve along the flux tube. The forward modeling code uses these quantities to synthesize spectra and broad-~/~narrow-band emission and folds the results through the appropriate wavelength resolved response functions. The code also transforms between numerical and instrumental spatial and temporal resolutions such that direct comparisons between model predictions and real observations can be made. In the present work we assume 0.33\arcsec pixels for \iris\ and a cadence of 5~s.

The set of element abundances provided by \cite{Asplund2009} have been used for the numerical and observational aspects of this investigation. The abundances of the elements of interest to us are photospheric. Oxygen is a high first ionization potential (FIP) element, whereas sulphur lies at the boundary between low- and high-FIP elements, though its abundance is consistently closer to photospheric. We have examined the relative abundances in different studies and found that the S/O ratio remains similar. The ratio used to produce the theoretical intensity ratio curves in Figure~\ref{fig_obs5} is 0.027 and other ratios all lie with 15-20\% of this value. The one outlier appears to be from \cite{Schmelz2012} who measure a ratio of 0.041, a factor of 1.5 greater, which may be attributable to choosing a value for the oxygen abundance at the very lowest end of the range plotted in their Figure~2. Furthermore, we did not find significant variation in the S/O ratio from one transition region feature to the next. In summary, we are satisfied that the choice of element abundances should not impact the generality of our findings in this work.

In Section~\ref{modelHydro} the importance of spatially resolving the transition region in numerical models was discussed with particular reference to accurately capturing its emission. In a poorly resolved transition region, where there are too few grid cells to capture the temperature structure, the peak population temperatures of the emitting ions can be missed. Consider an extreme example: in a hot loop with an essentially discontinuous temperature jump from $10^4$~K to $10^6$ between two grid cells at one foot-point, with no grid cells containing intermediate temperature plasma in the range that \oiv\ ($\log_{10}(T_e\textnormal{[K]})=5.1$), \siiv\ and \siv\ ($\log_{10}(T_e\textnormal{[K]})=4.9$) form, then no equilibrium emission would be predicted in the wavelength range to which \iris\ is sensitive. Furthermore, a non-equilbrium ionization calculation would also be inaccurate because ionization and recombination rates are temperature dependent. In less extreme cases of under-resolution the ion populations and the variability in the transition region emission are most likely to be underestimated in response to nanoflare heating.

In the set of numerical experiments described here the smallest grid cell is 244~m in the field-aligned direction. At the times when loop coronae reach their highest temperatures the greatest demands are placed on the grid resolution, because the transition region must steepen to handle the incoming conducted heat flux. In the weak-heating case the most powerful energy release onsets at 9589~s (Table~\ref{table3}) and reaches its peak one minute later at 9649~s, at which time the loop is expected to be hottest. There are grid cells containing plasma of temperature within 5\% of the peak population temperature of \oiv\ and within 25\% of the peak population temperature of \siiv\ and \siv. In the strong-heating case the most powerful nanoflare begins at 7236~s and peaks at 7296~s. There are grid cells containing plasma of temperature within 48\% of the peak population temperature of \oiv\ and within 17\% of the peak population temperature of \siiv\ and \siv. Note that these are the most extreme cases encountered during the numerical experiments, but they do serve to highlight the enormous difficulties that the demands associated with adequately resolving the transition region place on numerical codes.

\section{Observational Data}
\label{observations}

This work is focused on the effects due to non-equilibrium ionization and density-dependence on some of the \iris\ lines commonly used to perform spectroscopic diagnostics on transition region plasma. We are particularly interested in the \siv\ and \oiv\ lines in the range $1399-1406$~\AA, which provide diagnostics that are discussed in detail by \citep{Polito2016}. Given their popularity and diagnostic power, it is prudent to assess their sensitivity to the aforementioned processes. Since the \siv\ and \oiv\ lines are significantly weaker than the strong \siiv\ (1402.77~\AA~and 1393.78~\AA) lines in the \iris\ spectral range, we selected active region observations with long exposures to improve the signal-to-noise ratio. 

We present four active region observations carried out by \iris\ in September 2016. We analyze observations of active region (AR) 12593 (2016-09-21 starting near 00:14~UT and 2016-09-22 starting near 05:55~UT), AR 12595 (2016-09-21 starting near 06:21~UT), and AR 12597 (2016-09-27 starting near 22:22~UT). All selected active region datasets were obtained with the same observing sequence (\iris\ OBSID 3690215148) which uses 32-step sparse rasters (when "sparse" rastering the slit step is 1\arcsec), 60~s exposure times (for the spectra; the slit-jaw images have 8~s exposure times), and lossless compression. Slit-jaw images (SJI) were obtained while alternating the 1330~\AA, 1400~\AA, and 2796~\AA~passbands, yielding a cadence of $\sim 245$~s in each passband. The field of view (FOV) of the SJI is $\sim 122$\arcsec$\times 122$\arcsec\ and the FOV of the rasters is $\sim 31$\arcsec$\times 122$\arcsec. The datasets are re-binned by a factor of 4 yielding a spatial resolution of $\sim 0.665$\arcsec and $\Delta \lambda \sim 0.05$~\AA~for the FUV spectra. We use \iris\ calibrated level 2 data, which have been processed for dark current, flat field, and geometrical corrections \citep{DePontieu2014}.

\begin{figure}
	\plotone{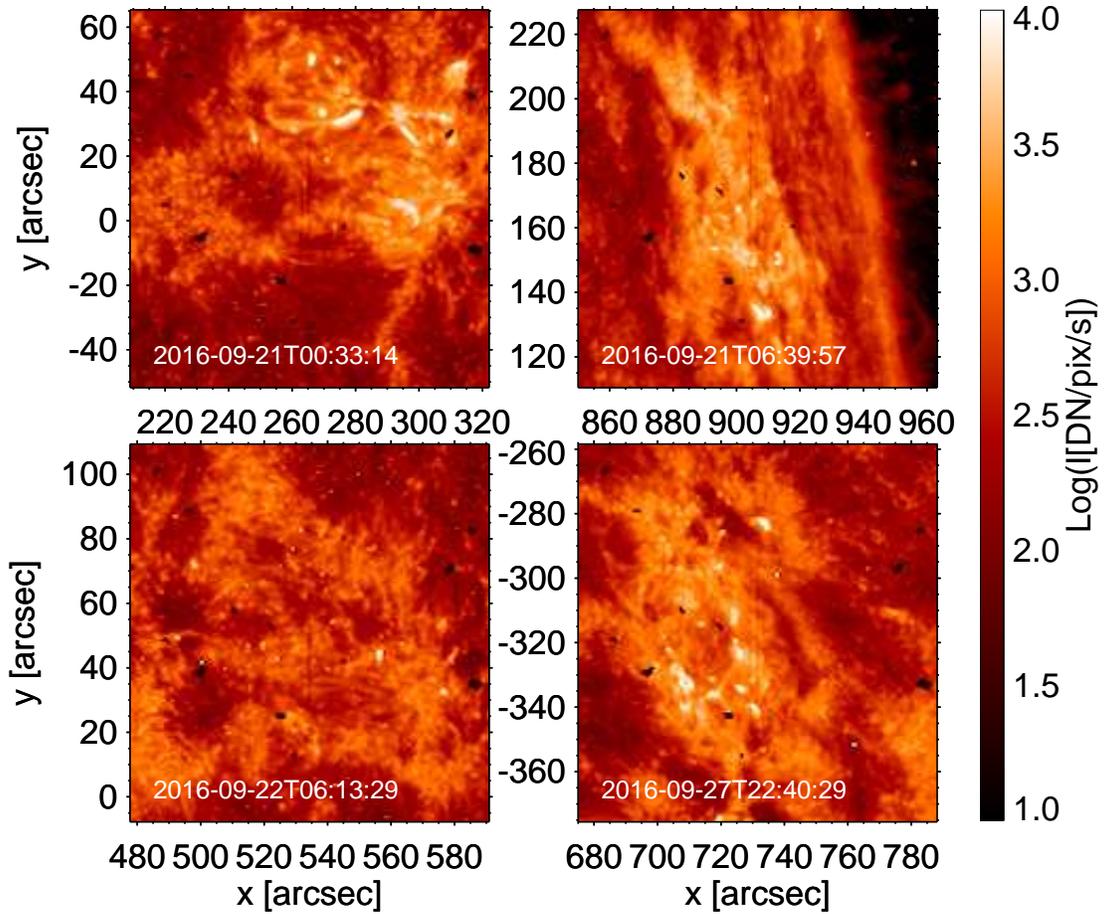}
	\caption{\iris\ SJI 1400~\AA~observations (dominated by transition region lines typically emitted by plasma at $\log_{10}(T_e[$K$])\sim4.7-5.2$) of AR 12593 (left column), AR 12595 (top right), and AR 12597 (bottom right). The slit is faintly visible as a vertical line near the center of each image.}
	\label{fig_obs1}
\end{figure}

In order to investigate the relationship between the transition region emission observed by \iris\ and the overlying corona, we use \aia\ observations. \aia\ observes the full Sun at high spatial resolution and temporal cadence ($\sim 0.6$~\arcsec and $\sim 12$~s, respectively) in several narrow extreme ultraviolet (EUV) channels, which sample the corona (and transition region) across a broad temperature range \citep{Boerner2012,Boerner2014}. We used \aia\ level 1.5 data, which is processed for the removal of bad-pixels, despiked, flat-fielded, and image registered (coalignment among the channels, with adjustment of the roll angle and plate scales).  

\begin{figure}
	\centering	
	\includegraphics[width=8cm]{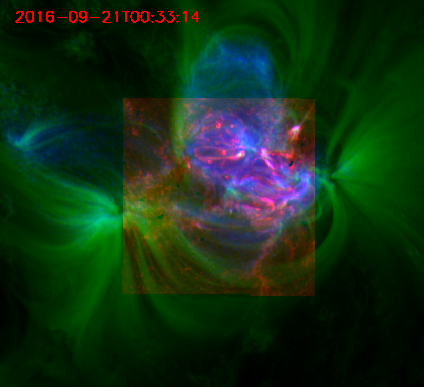}
	\includegraphics[width=8cm]{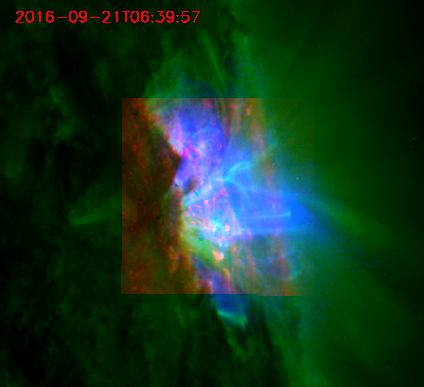}
	\includegraphics[width=8cm]{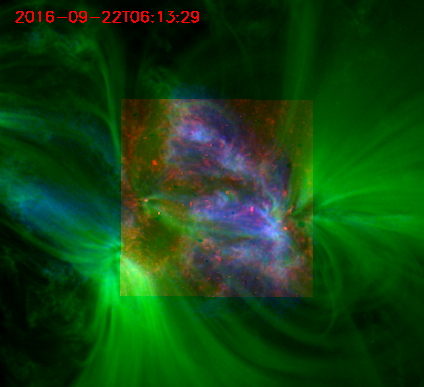}
	\includegraphics[width=8cm]{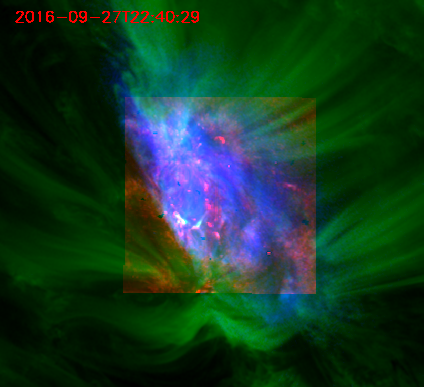}
	\caption{Three color images showing emission in different passbands~/~channels for the four active region datasets: \iris\ SJI 1400~\AA~dominated by transition region lines typically emitted by plasma at $\log_{10}(T_e[$K$])\sim4.7-5.2$ (red); \aia\ 171~\AA~dominated by \feix\ emission at $\log_{10}(T_e[$K$])\sim5.9$ (green); and \aia\ 335~\AA~ which includes a strong \fexvi\ line emitted at $\log_{10}(T_e[$K$])\sim6.5$ (blue).}
	\label{fig_obs2}
\end{figure}

In Figure~\ref{fig_obs1} we show four \iris\ SJI 1400~\AA~active region observations (the 1400 SJIs are obtained with exposure times of 8~s). The 1400~\AA~\iris\ passband is generally dominated by transition region emission and in particular the strong \siiv\ 1402.77~\AA~and 1393.78~\AA~lines \citep{DePontieu2014}. In Figure~\ref{fig_obs2} we have superimposed \aia\ images in the 171~\AA~channel, dominated by $<1$~MK plasma, and in the 335~\AA~channel, dominated by $\sim 3$~MK plasma, on the \iris\ 1400~\AA~SJIs. The brightest active region features in these \aia\ channels are generally the long, cool loops at the periphery of the active regions in the 171~\AA channel, and the hot emission from the active region cores in the 335~\AA channel. The \aia\ images indicate that the selected active regions are characterized by different morphologies, activity levels, and inclinations with respect to the line-of-sight. We note that the datasets of 2016-09-21 at 00:00~UT and 2016-09-22 at 06:00~UT observe the same active region (AR 12593) 30~hrs apart, and show significant decay of that region. The \aia\ data have been aligned with the \iris\ SJI data by applying a standard cross-correlation routine (tr\_get\_disp.pro, which is part of the IDL SolarSoftware package) to \aia\ 1600~\AA~and \iris\ 1400~\AA~SJI images, which generally show similar morphology.

The intensities of the \siv\ and \oiv\ transition region lines were measured by a Gaussian fit to the \iris\ FUV spectra. As noted by \cite{Polito2016} and \cite{Dudik2017} these lines sometimes show departures from Gaussian shapes and can be well-fitted by a Lorentzian or double Gaussian function. We used a double Gaussian to fit each line and cross-checked each fit by performing an additional single Gaussian fit, finding that the line intensities are consistent with the double Gaussian method to within $\sim 15-20$\%. We measured the \oiv\ lines at 1399.780~\AA, 1401.157~\AA, and 1404.806~\AA, and the \siv\ line at 1406.016~\AA. We used the measured \oiv\ line intensity ratios (1399~/~1401~\AA~and 1401~/~1404~\AA) for plasma density diagnostics. We note that the 1404~\AA~line is blended with a \siv\ line that has theoretical rest wavelength (1404.808~\AA) very close to the \oiv\ line (1404.806~\AA). The \siv\ contribution to the 1404~\AA~blend can be estimated using the measured intensity of the \siv\ 1406~\AA~line: the \siv\ 1404~/~1406~\AA~intensity ratio, predicted by \chianti, varies between $\sim 0.21$ at $\log_{10}(T_e[$K$])=4.8$ and $\sim 0.19$ at $\log_{10}(T_e[$K$])=5.2$ for densities up to at least $\log_{10}(n_e[$cm$^{-3}])\sim11$ (as noted by \citealt{Polito2016}, the \siv\ line ratio can be used for density diagnostic in flares because of its sensitivity to especially high densities). We note that the \oiv\ density diagnostics give us typical densities between $10^{10}$ and $10^{11}$~cm$^{-3}$, and we estimate the \siv\ 1404~\AA~line intensity with 0.2 as scaling factor from the measured \siv\ 1406~\AA~line intensity. Furthermore, the 1404 \oiv\ line is only used to derive a density from the \oiv\ 1401~/~1404~\AA\ ratio and then used to cross-check the densities derived from the 1399~/~1401~\AA\ ratio; we use the latter ratio as the main diagnostic of plasma density.

\begin{figure}
	\plotone{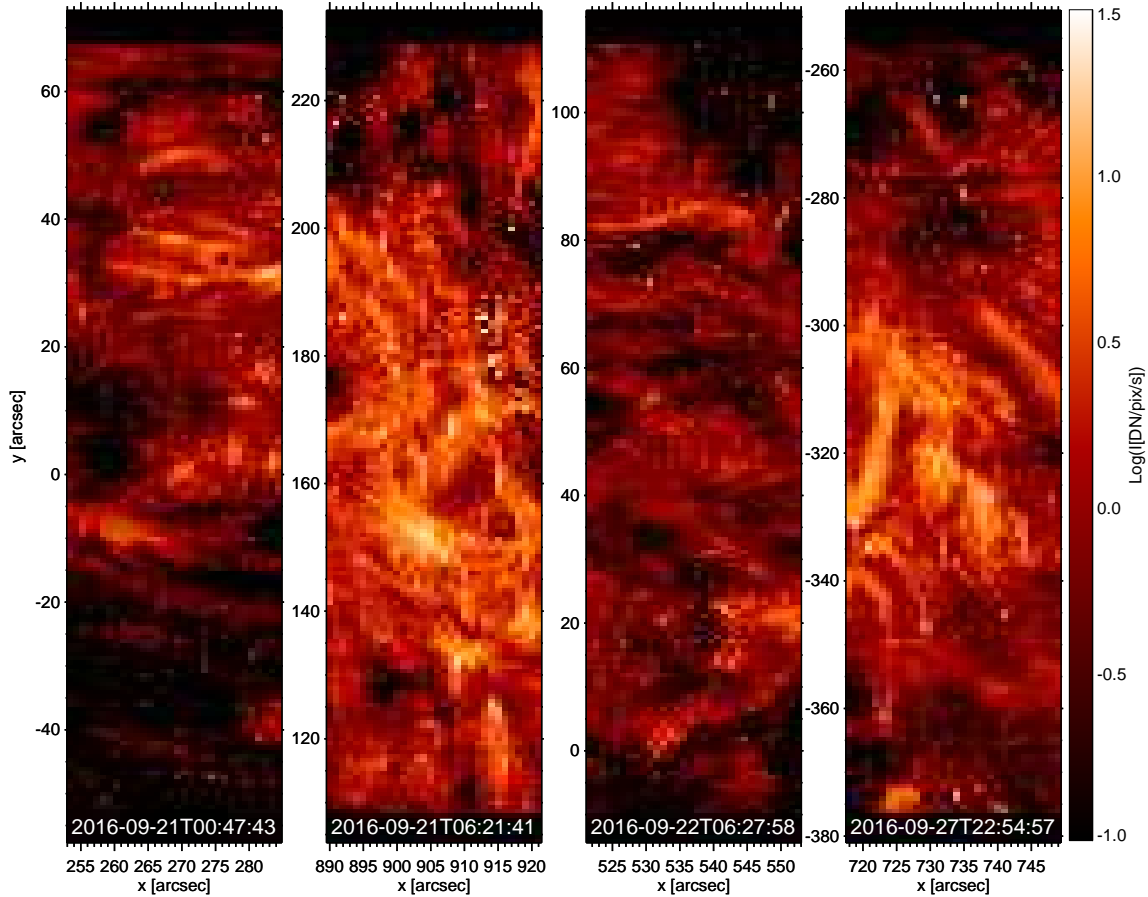}
	\caption{Intensity maps of the \oiv\ 1401~\AA~transition region line, in the \iris\ raster FOV, for the active region observations shown in Figures~\ref{fig_obs1} and \ref{fig_obs2}. The color bar shows the corresponding intensity values in units of DN/pixel/s, where the pixel is the original \iris\ pixel (the measured data numbers are divided by the exposure time and by 16, given the binning $\times$4 in both spatial and spectral dimensions).}
	\label{fig_obs3}
\end{figure}

In Figure~\ref{fig_obs3} we plot intensity maps for the \oiv\ 1401~\AA~line in the FOV of the \iris\ rasters. We note that the intensity units used for the plots in Figure~\ref{fig_obs3} are directly comparable to the predicted values, taking the numerical results as input to the forward model, as shown in Figure~\ref{fig2} later on (i.e., DN per "unbinned" \iris\ pixel per second). The intensity maps for the \siv\ 1406~\AA~line have very similar morphology to these \oiv\ 1401~\AA~maps shown in Figure~\ref{fig_obs3}.

\begin{figure}
	\plotone{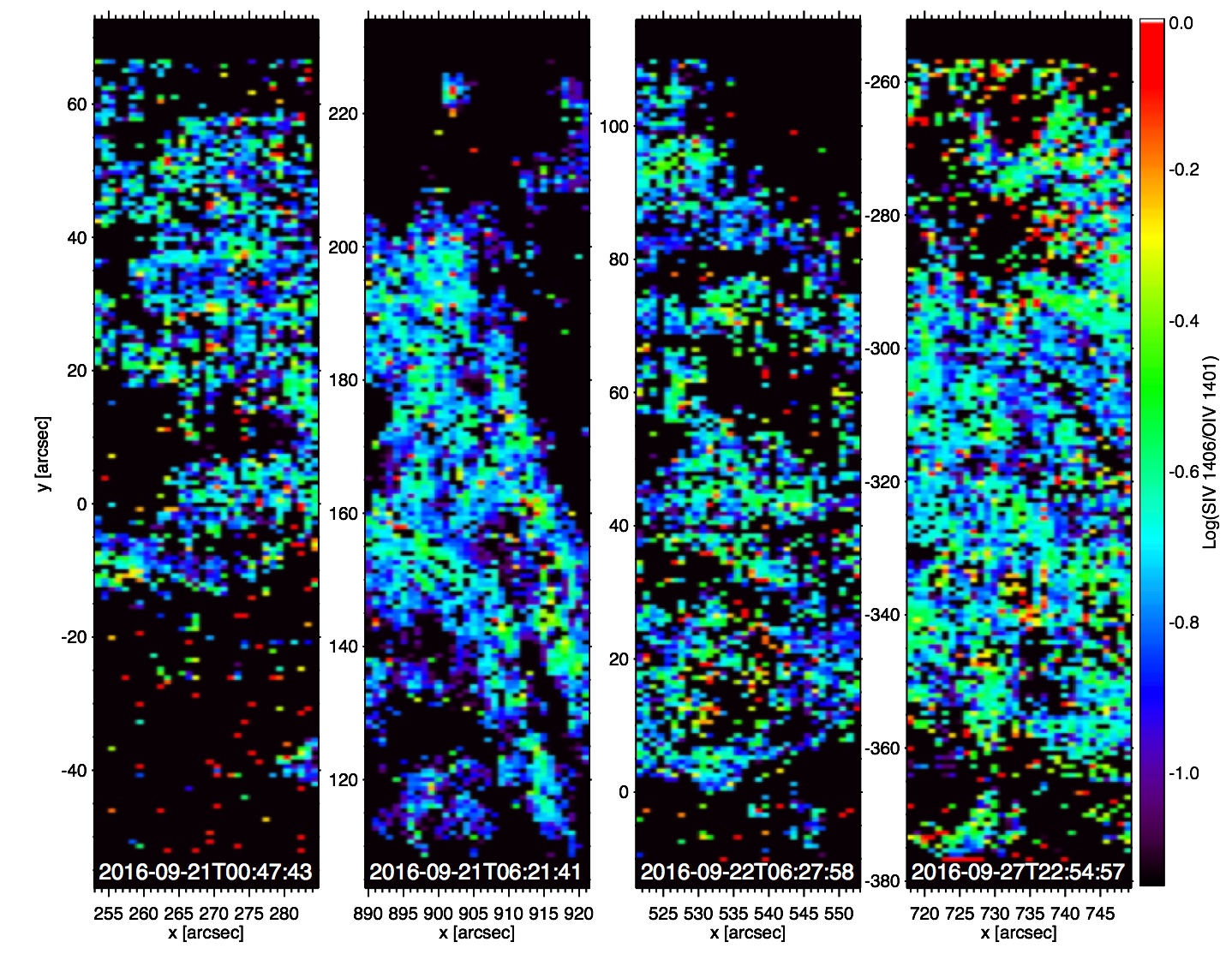}
	\caption{Maps of the \siv\ 1406~\AA~to \oiv\ 1401~\AA~ratio, in the \iris\ raster FOV, for the active region observations shown in Figures~\ref{fig_obs1} to \ref{fig_obs3}.}
	\label{fig_obs4}
\end{figure}

\begin{figure}
	\plotone{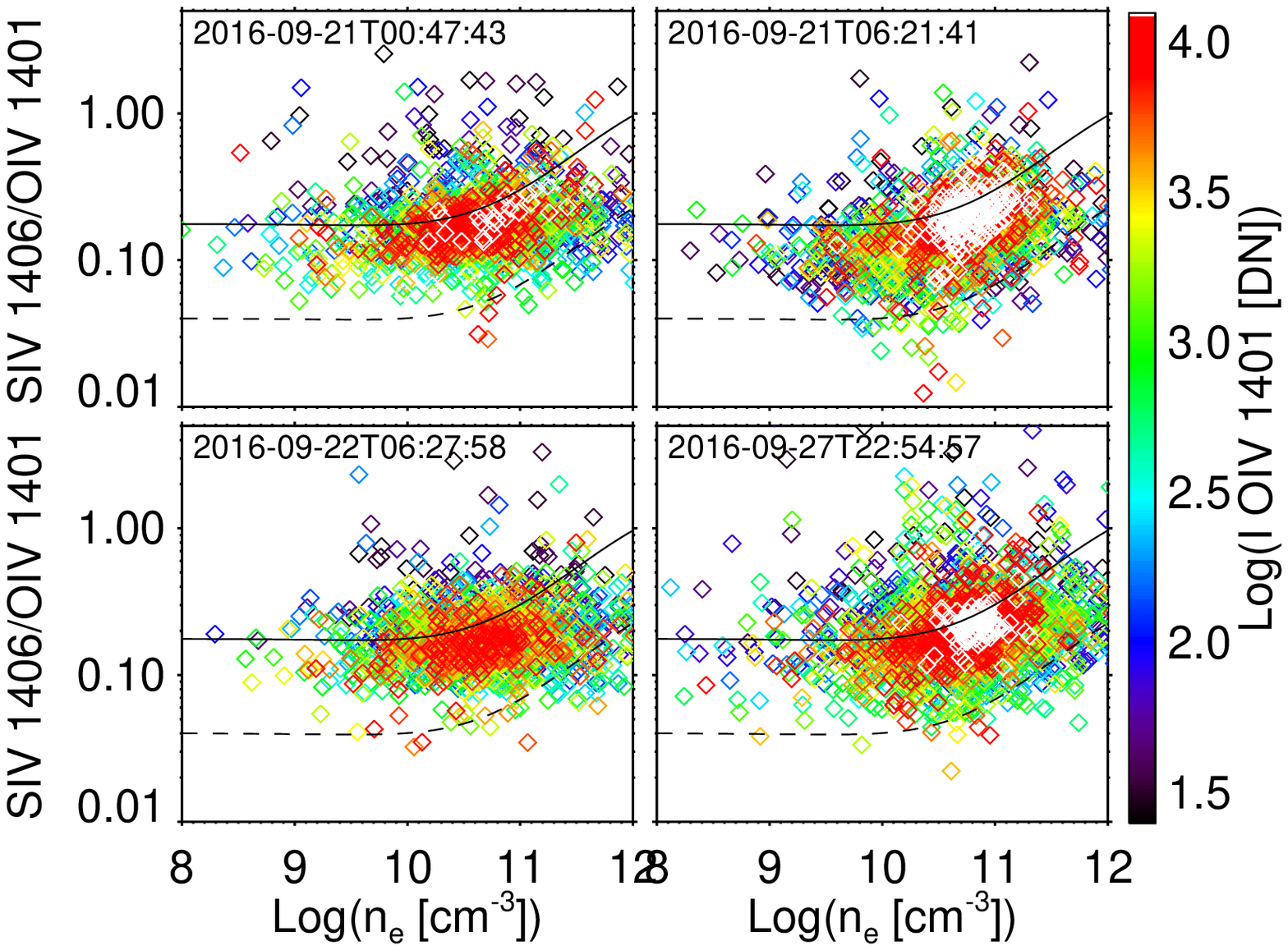}
	\caption{Scatter plots of the \siv/\oiv\ line ratio (also shown in Figure~\ref{fig_obs4} as spatial maps), as a function of the electron density, for the active region observations presented in Figures~\ref{fig_obs1} to \ref{fig_obs3}. The solid~/~dashed lines correspond to the theoretical ratios from \chianti\ for $\log_{10}(T_e[$K$])=5$ (solid), which is the peak of the emissivity function of the \siv\ line, and for $\log_{10}(T_e[$K$])=5.15$ (dashed), which is the peak of the emissivity function of the \oiv\ line. The colors correspond to the line intensity so that signal-to-noise effects can be investigated. The white data points are saturated values ($>10^4$~DN).}
	\label{fig_obs5}
\end{figure}

The introductory discussion in Section~\ref{introduction} provided details concerning previous work that studied the effects of non-equilbrium processes on the formation of emission lines in the \iris\ passbands, focusing in particular on \siiv/\oiv\ ratios as a diagnostic of departures from equilibrium \citep{Olluri2013b,Martinez2016,Dzifcakova2018}. We are concerned with determining which atomic processes are of most import to the formation of \siv\ lines and we will use \siv/\oiv\ ratios as a diagnostic tool to pursue this investigation. In Figure~\ref{fig_obs4}, we plot maps of the ratio of the \siv\ 1406~\AA~line to the \oiv\ 1401~\AA~line for the four active region datasets. In Figure~\ref{fig_obs5} we show scatter plots of these \siv\ to \oiv\ line ratios as a function of the electron density (derived using the \oiv\ 1399~/~1401~\AA~line ratio).

The plots in Figure~\ref{fig_obs5} highlight a few key properties of the relationship between the \siv/\oiv\ ratio and the electron density. The bulk of the plasma appears to be characterized by typical electron densities of $\log_{10}(n_e[$cm$^{-3}])\sim10-11$. The measured \siv/\oiv\ ratios tend to cluster around the curve corresponding to the lower formation temperature of \siv, rather than some intermediate temperature between the \siv\ and \oiv\ formation temperatures. The data points at greater \siv/\oiv\ values, corresponding to lower temperatures, exhibit a tendency toward lower intensities, which perhaps indicates a signal-to-noise effect. The plots of Fig.~\ref{fig_obs4} and ~\ref{fig_obs5} do not seem to indicate any obvious relationship between measured \siv/\oiv\ ratios and the activity level of the active region.

\section{Numerical Results}
\label{results}

\begin{figure}
	\plotone{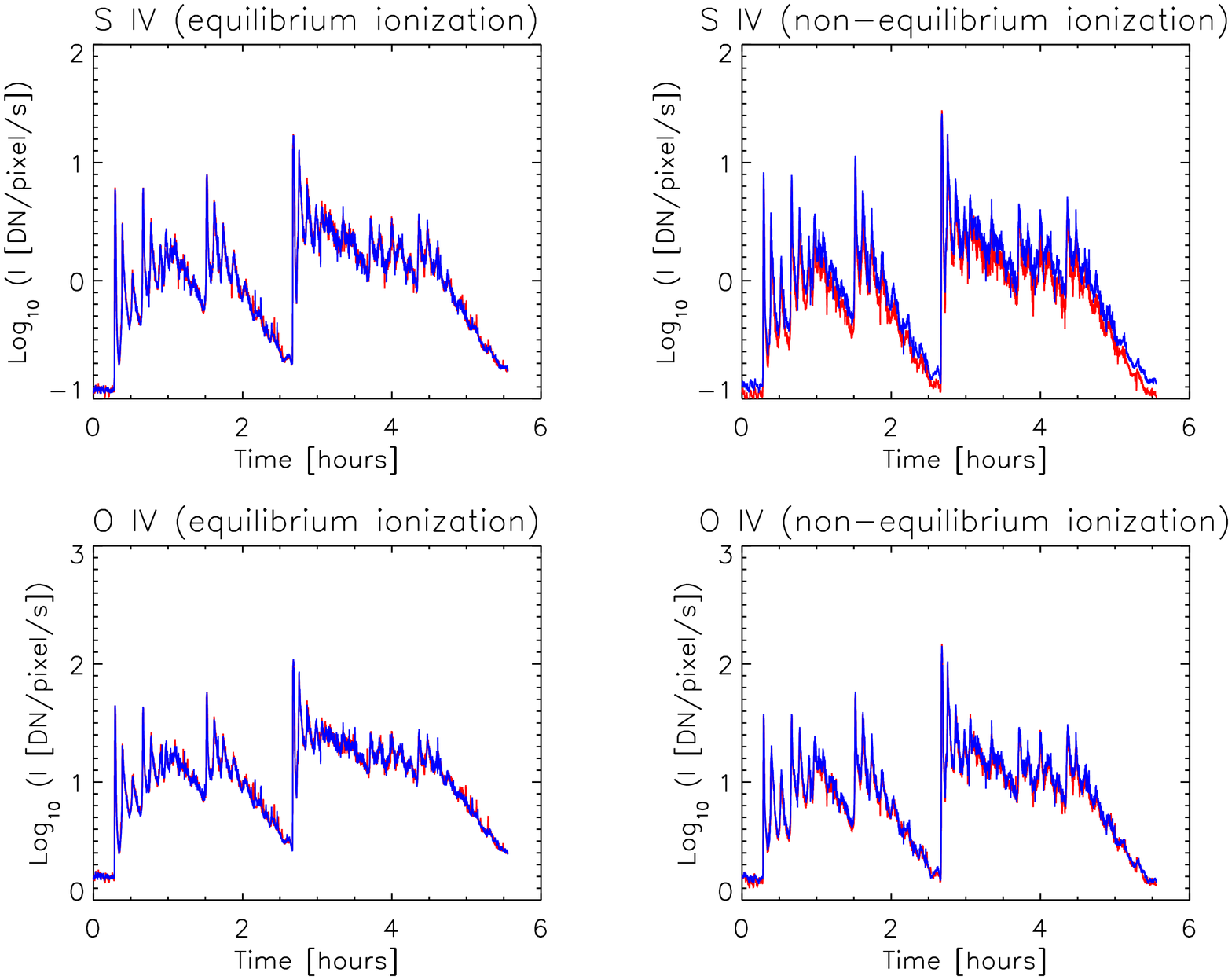}
	\caption{The predicted line-of-sight integrated intensities of the \oiv\ (1401.16~\AA) and \siv\ (1406.06~\AA) emission lines in the weak-heating case. The red curves show the intensities calculated in the low-density limit and the blue curves when density-dependence is accounted for. The intensities have been calculated for an equilibrium (non-equilibrium) ion population in the left (right) column.}
	\label{fig2}
\end{figure}

Figure~\ref{fig2} shows predicted line-of-sight integrated intensities, in instrument units, for two emission lines in the \iris\ wavelength range calculated for the weak heating case. The curves show results in the low density limit and when density dependence is accounted for in the formation of the ion population under equilibrium and non-equilibrium conditions. Ionization and recombination rates provided by the {\it ADAS} database were used for this part of the investigation. The emission exhibits significant variability in each case and the onset times of periods of particular variability are well-correlated with the heating events shown in Figure~\ref{fig1}. When the ion population is treated in equilibrium, density dependence has no clear influence on the intensities of the \oiv\ or \siv\ lines. In non-equilibrium one can see, by eye, that the variability is somewhat enhanced in both cases, but also the intensity of the \siv\ line is systematically increased in the density dependent case. \siv\ forms at a slightly lower temperature, and hence higher density in the stratified solar atmosphere, than \oiv. During nanoflare heating the coronal pressure increases and pushes the transition region deeper into the solar atmosphere; the density-dependent quenching of dielectronic recombination from \siv\ to S~III increases the lifetime of \siv\ as it is transported into a higher density region and continues to radiate in proportion to $n_e^2$. Hence, the intensity of \siv\ emission increases. It is important to note that one must account for non-equilibrium ionization in order to observe the effect of density dependence; these processes are strongly coupled in this instance.

\begin{figure}
	\plottwo{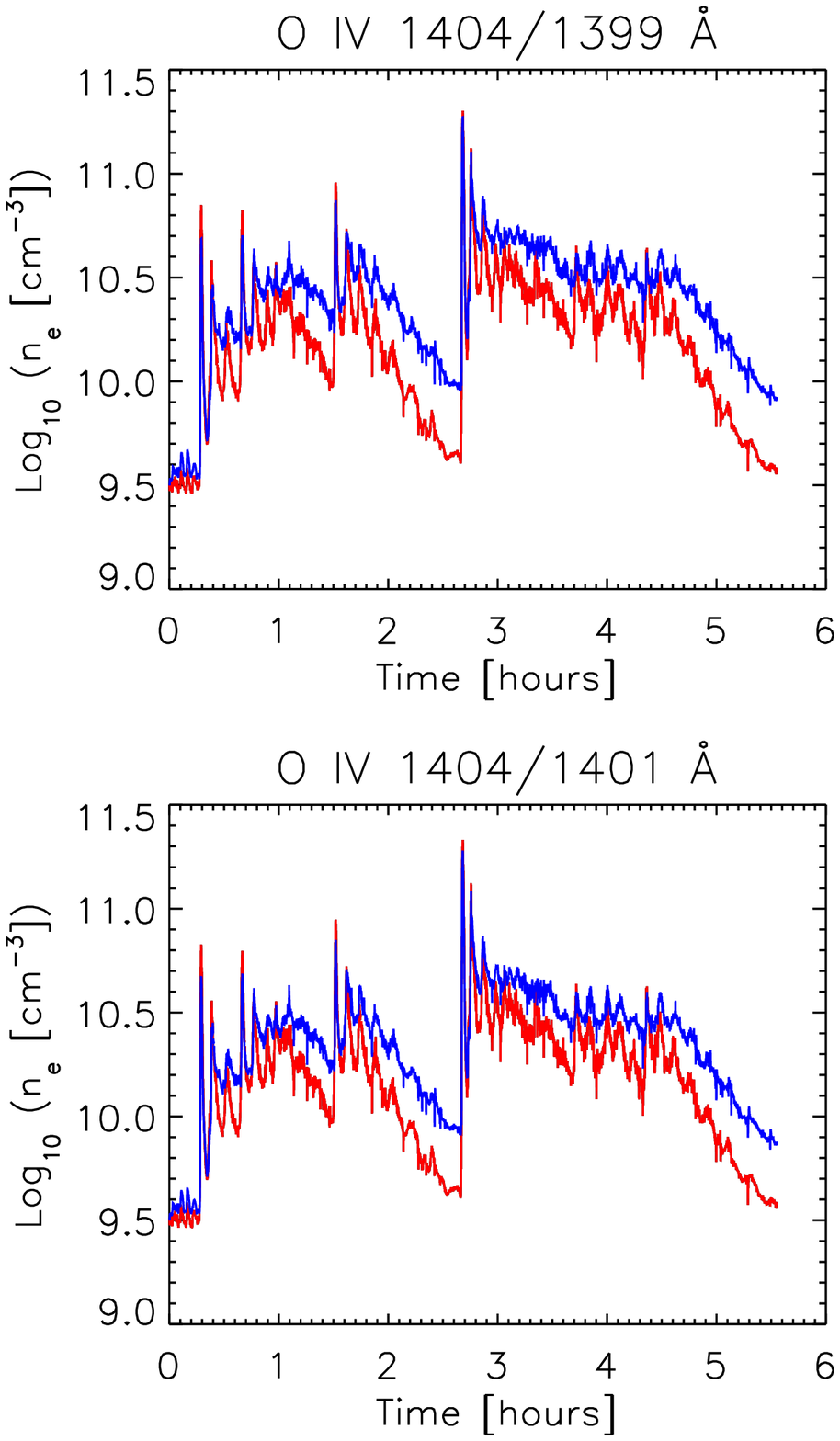}{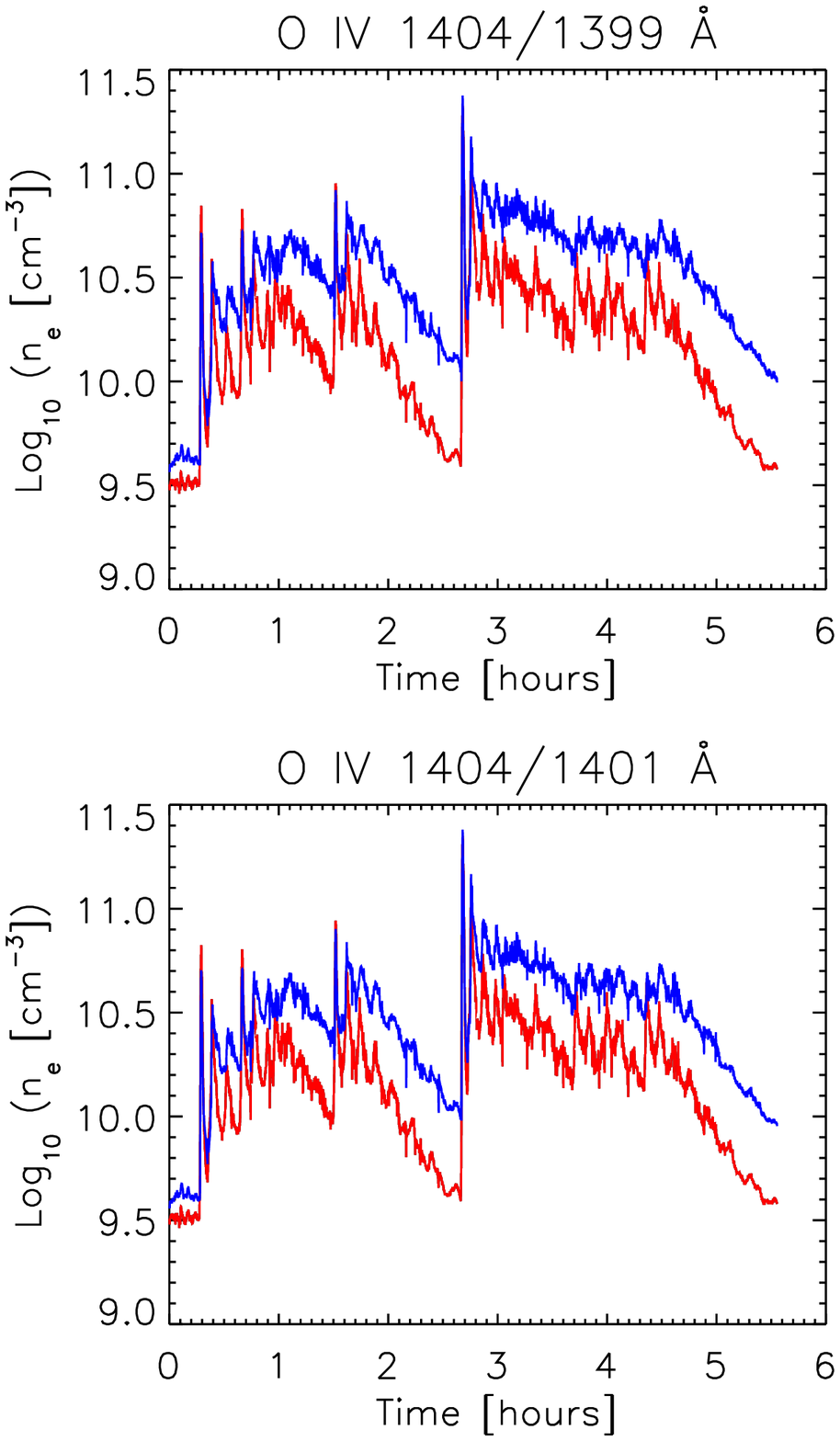}
	\plottwo{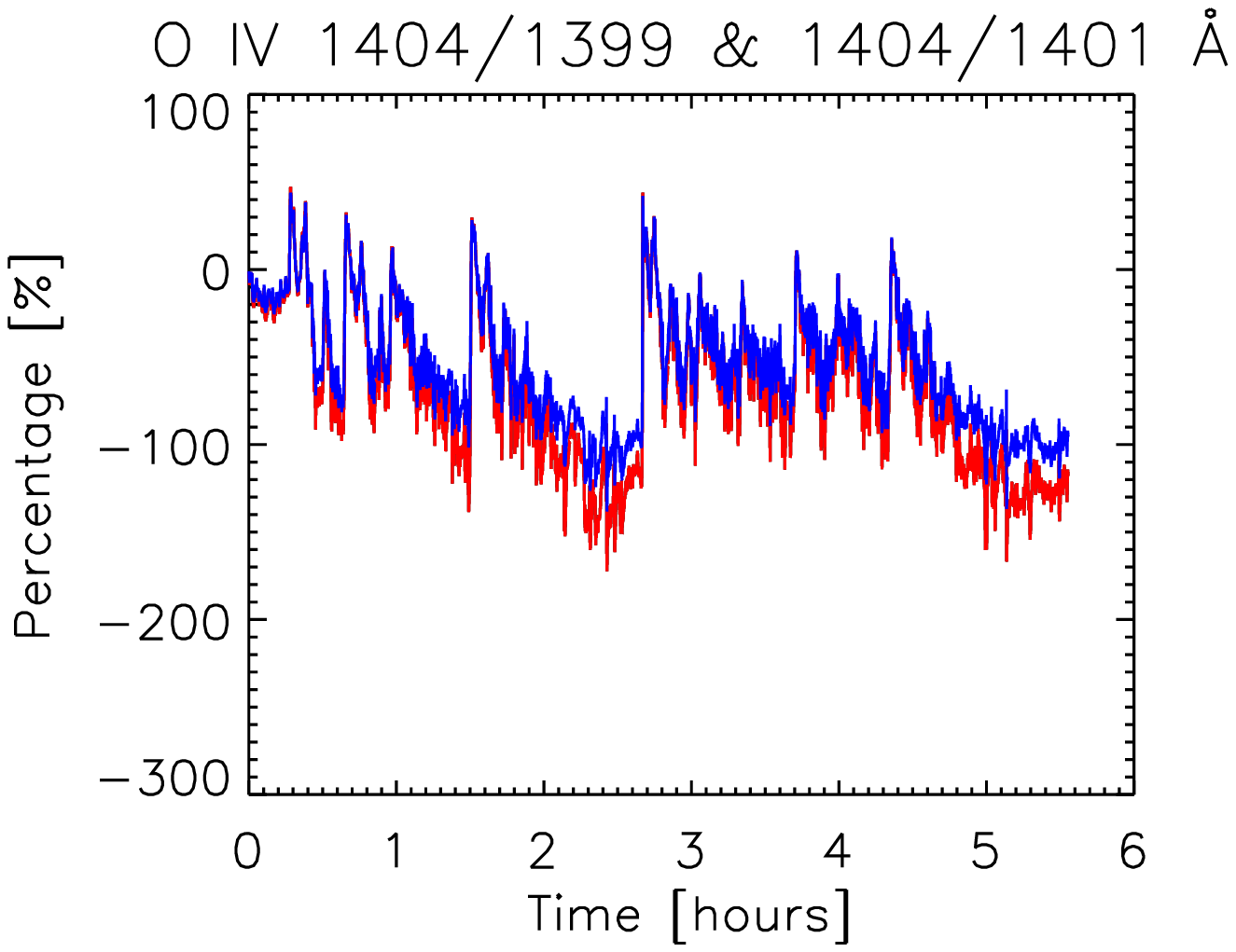}{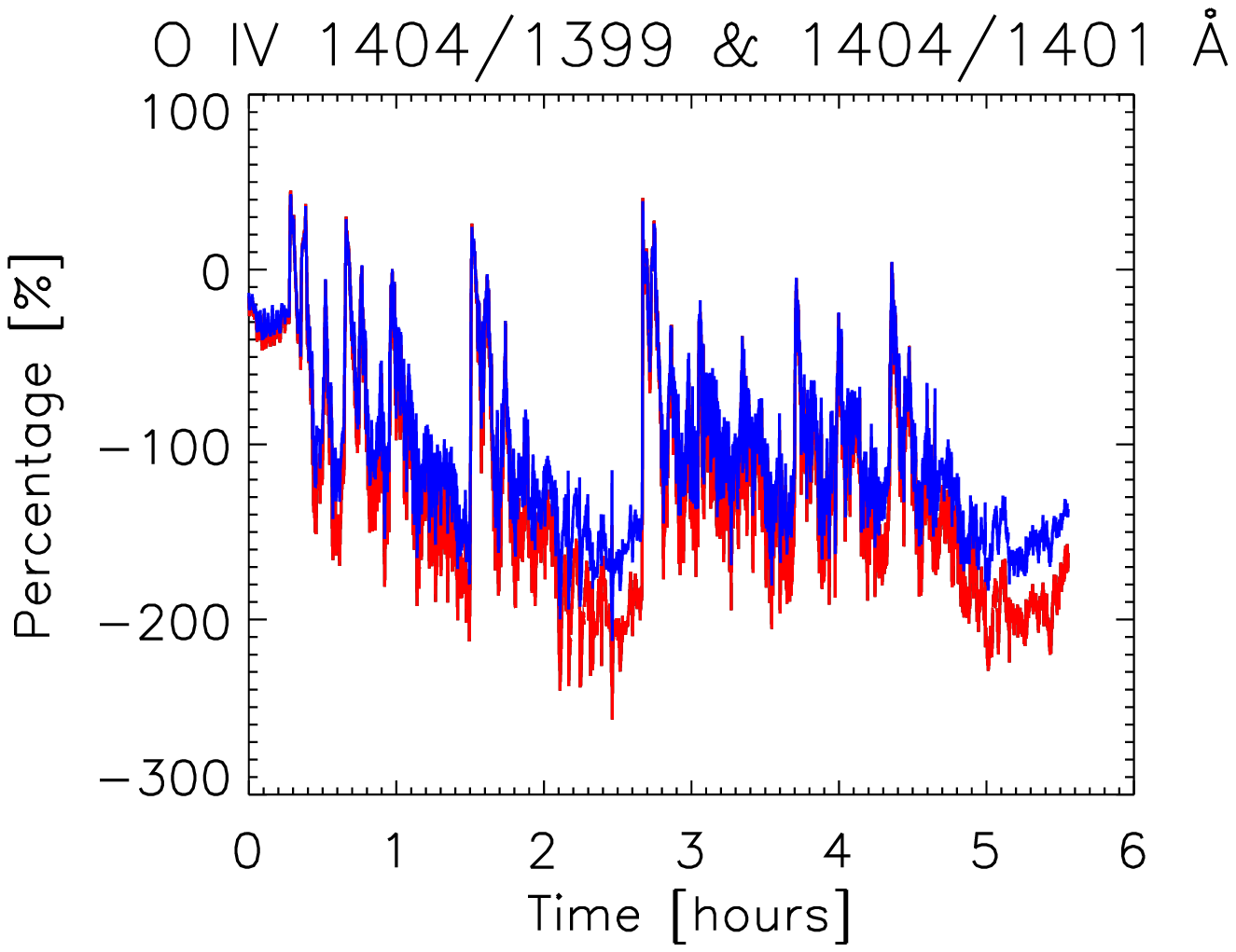}
	\caption{The predicted electron density, calculated from two sets of line-of-sight integrated, density-sensitive, line ratios, in the weak-heating case. The results presented in the left column were calculated in the low-density limit. The results in the right column were calculated when accounting for density dependence. The top (middle) row shows the predicted density with the \oiv\ 1404~/~1399 (1404~/~1401)~\AA~line pair. In each of these plots the red (blue) curve is calculated for an equilibrium (non-equilibrium) ion population. The bottom row shows the difference between each line pair calculated for equilibrium and non-equilibrium ion populations. The red (blue) curve corresponds to the 1399.78 (1401.16)~\AA~pair.}
	\label{fig3}
\end{figure}

Based on Figure~\ref{fig2} one might assert that density dependence has relatively little effect on the \oiv\ emission in the \iris\ spectral window, even in the presence of a non-equilibrium ionization state. In fact the effect is subtle and simply comparing line intensities does little to reveal it, but an example using density sensitive \oiv\ line pairs can provide a stark illustration of its importance for density measurements obtained via common spectroscopic diagnostics. Figure~\ref{fig3} shows the predicted electron density calculated from two sets of line-of-sight integrated, density-sensitive line ratios: \oiv\ 1404~/~1399~\AA; and \oiv\ 1404~/~1401~\AA. The results show that in every case, non-equilibrium ionization yields a greater estimate of the density than the corresponding equilibrium case. This is for the same reason discussed in the previous paragraph: the lifetime of the ion allows it to persist just long enough to emit at a higher density. Most crucially, however, in the density-dependent case (right column), the difference between the equilibrium and non-equilibrium ionization cases are enhanced due to the quenching of dielectronic recombination increasing the lifetime of the ion at higher densities. One can see this quantitatively in the bottom two plots of Figure~\ref{fig3}, which show the percentage difference between the predicted densities according to an equilibrium and non-equilibrium ion population, for each line pair. A negative percentage indicates an overestimate relative to the equilibrium value. It is evident from these plots that the 1399.78~\AA~ratio is more susceptible than the 1401.16~\AA~ratio to the combination of non-equilibrium ionization and density dependence. However, it is also important to realize that the densities corresponding to non-equilibrium ionization are measurements at the depth where the emission is strongest, which is not necessarily at the temperature of maximum ion population fraction in equilibrium.

\begin{figure}
	\plottwo{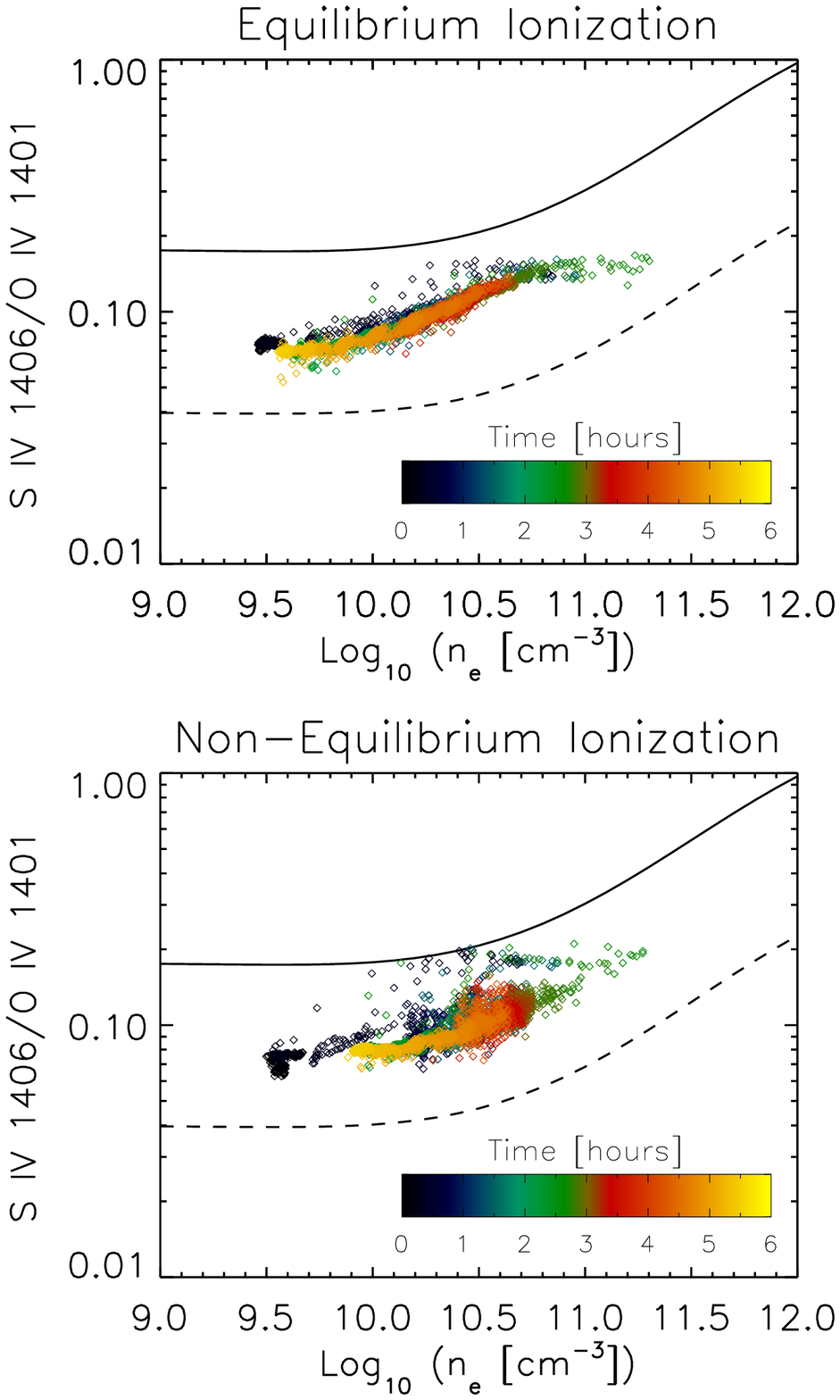}{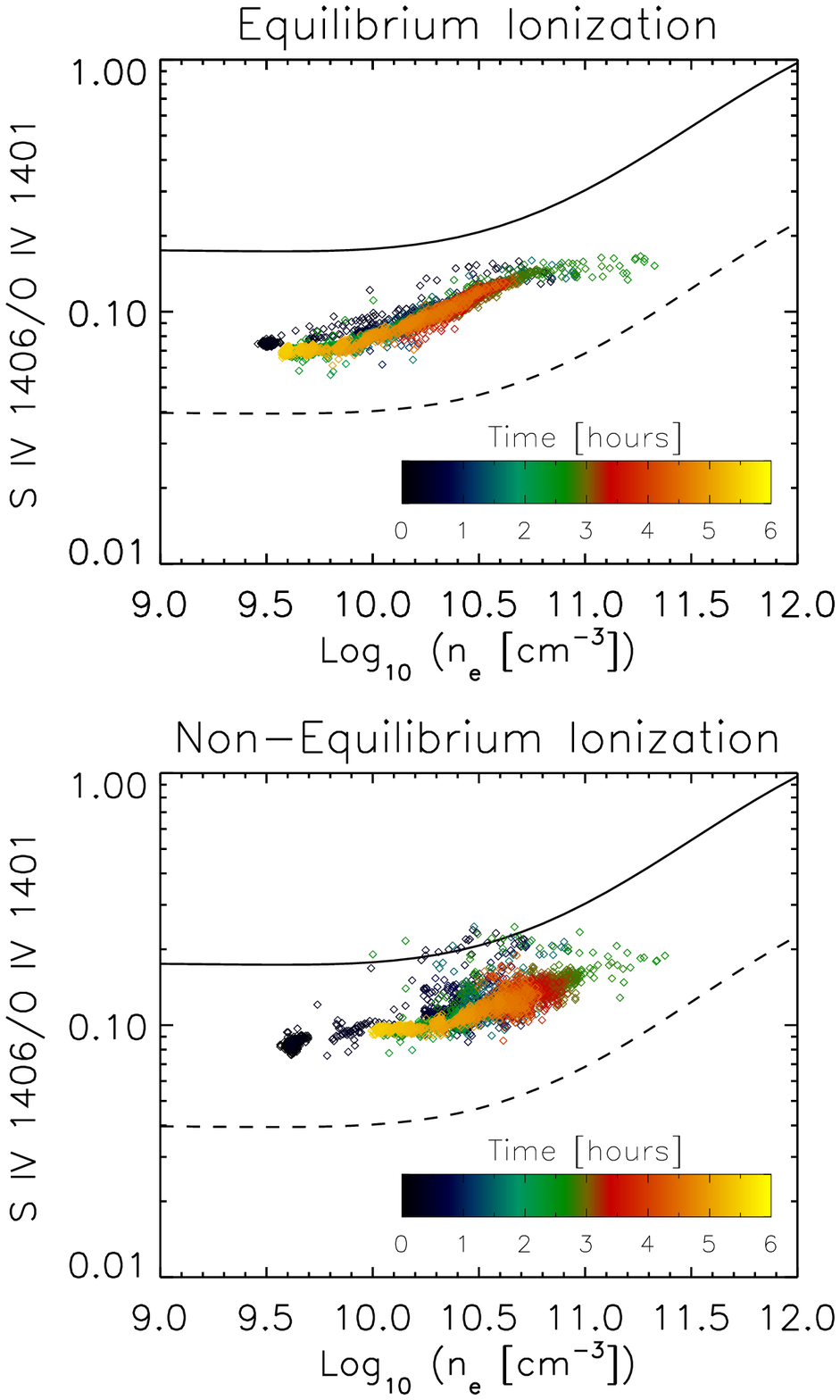}
	\caption{The predicted \siv\ to \oiv\ intensity ratio, for the emission lines shown in Figure~\ref{fig2} in the low-density limit (left column) and accounting for density dependence (right column), plotted as a function of electron density. The electron density was calculated using the predicted \oiv\ 1404~/~1399~\AA~ratio (Figure~\ref{fig3}). The data points are color-coded according to the color bar. The dashed line is the \chianti-predicted ratio at the formation temperature of \oiv\ in equilibrium and the solid line is the predicted ratio at the formation temperature of \siv\ in equilibrium.}
	\label{fig4}
\end{figure}

Figure~\ref{fig4} shows a set of color-coded plots for a \siv\ (1406.06~\AA) and \oiv\ (1401.16~\AA) line ratio plotted as a function of number density in the weak-heating case. The reason for this choice of emission lines from these particular ions is that we expect their ionization timescales to be significantly different to each other, and plotting their intensity ratio then provides a measure for the strength of departures from ionization equilibrium in the emitting region. Ratios calculated for ion populations in equilibrium are plotted in the top two panels of Figure~\ref{fig4}, while the ratios for non-equilibrium populations are shown in the lower two panels. The plots are relatively straightforward to interpret. For example, Figure~\ref{fig1} shows a (relatively) strong heating event between 2.5 and 3 hours (at approx. 2 hours and 40 minutes) in the weak-heating case, which drove a steep enhancement in \siv\ and \oiv\ emission (Figure~\ref{fig2}), and a sharp density increase (Figure~\ref{fig3}). This time period corresponds to the green-colored data points in Figure~\ref{fig4}, which exhibit strong excursions to densities above $10^{11}$~cm$^{-3}$ consistent with what is shown in Figure~\ref{fig3}. When the ionization state is in equilibrium, the data points are expected to fall approximately halfway between the dashed and solid curves in Figure~\ref{fig4}, which are the \chianti-predicted ratios of the emission lines at the equilibrium formation temperatures of \oiv\ and \siv, respectively. As the ionization state departs from equilibrium the range of ratio values increases at a given density, and at particularly strong departures from equilibrium ionization the data points fall outside the area enclosed by the two curves. An example of this can be seen when accounting for density dependence, as shown in the right column of Figure~\ref{fig4}.

As expected, density dependence has little effect in the case of equilibrium ionization. However, in the non-equilibrium case several data points appear above the solid curve, indicating strong departures from the equilibrium state. Furthermore, we ascertain that these strong departures only occur during the first 3 hours of the model run (black to green data points). The later data points (red to yellow) fall within the area bounded by the \chianti\ curves. What is special or peculiar about the first 3 hours of the weak heating case? We can see from Figure~\ref{fig1} that it was characterized by stronger average coronal temperature peaks (6~MK), compared to later times (4~MK), that clearly increased the coronal pressure, and forced the transition region deeper into the atmosphere. This increased the density at which \siv\ and \oiv\ were formed, due to the combined effect of non-equilibrium ionization and density-dependent dielectronic recombination, which is evident in the sharp density enhancements in Figure~\ref{fig3}. In addition, since \siv\ forms at a lower temperature and higher density (in a stratified atmosphere) than \oiv, and the emission line intensity scales as $n_e^2$, this led to an increase in the \siv~/~\oiv\ ratio, as seen in Figure~\ref{fig4}. The enhanced \siv\ emission in non-equilibrium ionization, compared to equilibrium, can also be discerned from the peaks in Figure~\ref{fig3} during the first 2 hours of the weak heating case. The \oiv\ emission is similar in the equilibrium and non-equilibrium calculations. The reason for the reduced temperature (and hence pressure) peaks at later times, despite the similar magnitude of volumetric heating in each event, is that the corona becomes denser (Figure~\ref{fig1}) because its rate of cooling and draining is slow compared to the rate of ablation from the lower atmosphere, and so there is less energy per particle available to energize the coronal plasma and setup a strong pressure gradient.

\begin{figure}
	\centering
	\includegraphics[scale=0.60]{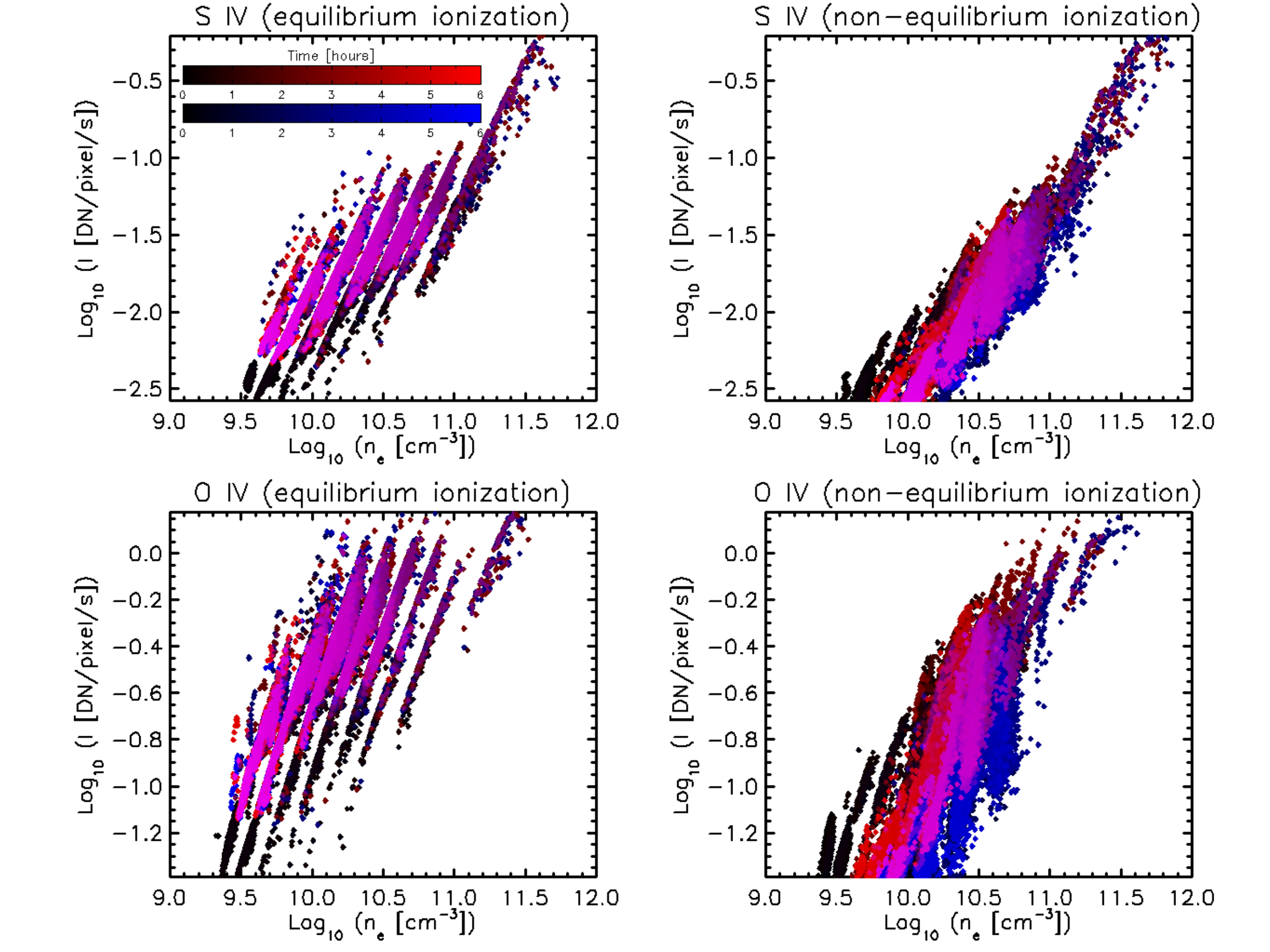}
	\\
	\includegraphics[scale=0.60]{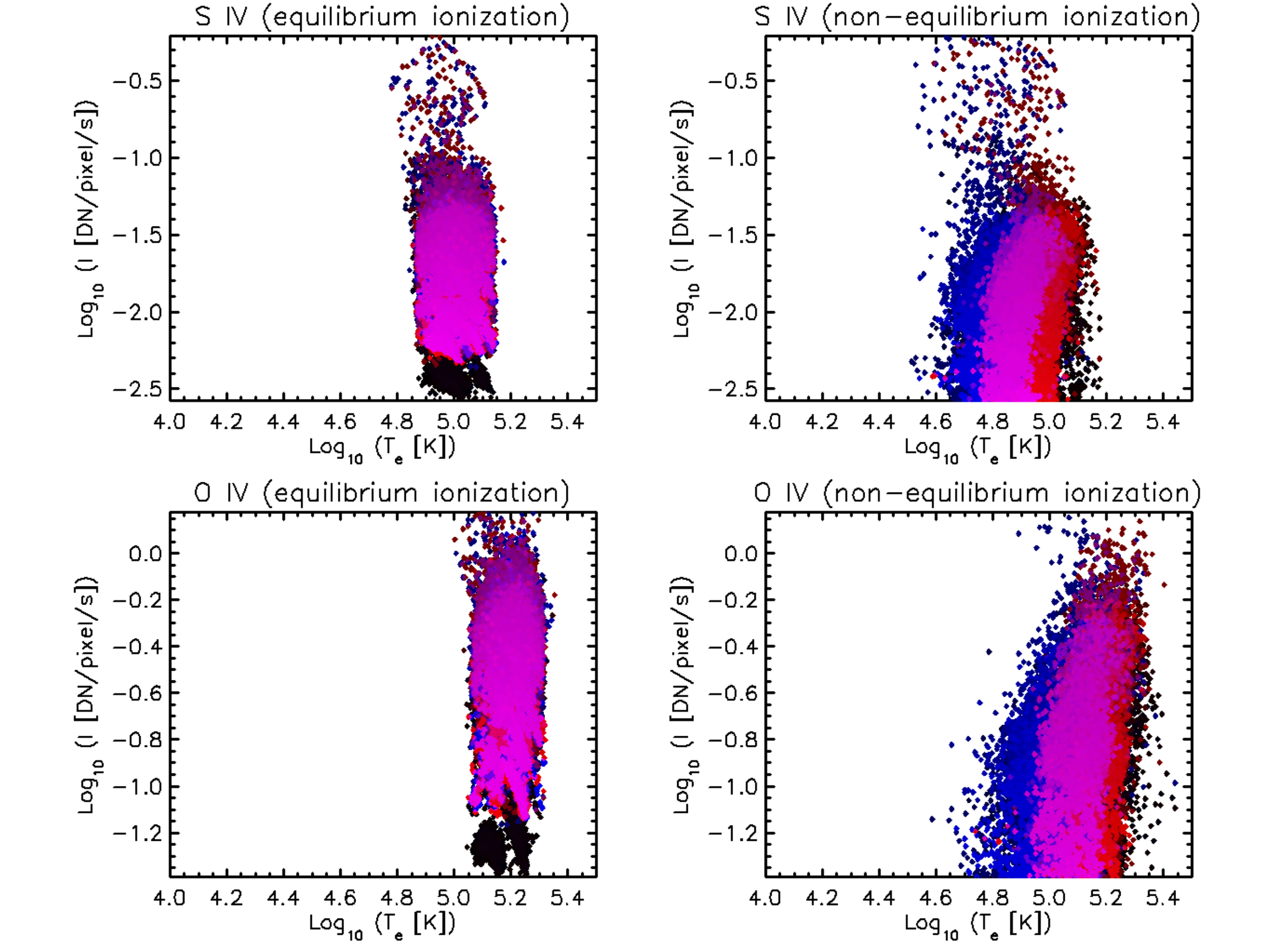}
	\caption{The predicted intensities for the emission lines shown in Figure~\ref{fig2} plotted as a function of density (upper four panels) and temperature (lower four panels). The red (blue) data points correspond to intensities calculated in the low-density (density-dependent) limit. Purple data points show the regions of overlap.}
	\label{fig5}
\end{figure}

Figure~\ref{fig5} shows the predicted distribution of densities and temperatures over which the \oiv\ and \siv\ lines emit in the weak-heating case. In the case of equilibrium ionization there is very strong overlap between the low-density limit and when density-dependence is accounted for in the dielectronic recombination rate. The non-equilibrium ionization case offers some revealing insights. Examining the intensity versus density column of Figure~\ref{fig5} one can observe a trend for the intensity to fall and shift towards higher densities as it does so. For example, the number of data points and the intensity at $n_e=10^{10}$~cm$^{-3}$ are significantly reduced compared to the corresponding equilibrium ionization prediction. Clearly, the combination of non-equilibrium ionization and quenching of dielectronic recombination has allowed these populations to penetrate deeper into the atmosphere. The reason for reduced intensities can be discerned from the intensity versus temperature column. While the warm parts of the distributions remain approximately fixed in temperature, the cool part extends significantly to lower temperatures: by almost 0.4~dex in the case of \oiv\ and 0.2~dex for \siv. The enhanced population at lower temperatures, despite the increase in density, explains the reduced intensities. However, the line-of-sight integrated intensities are predicted to be stronger as a consequence of impulsive heating (e.g. Figure~\ref{fig2}) and these findings suggest that this is due to an increased column depth of the emitting ions rather than a density effect alone.

\begin{figure}
	\plottwo{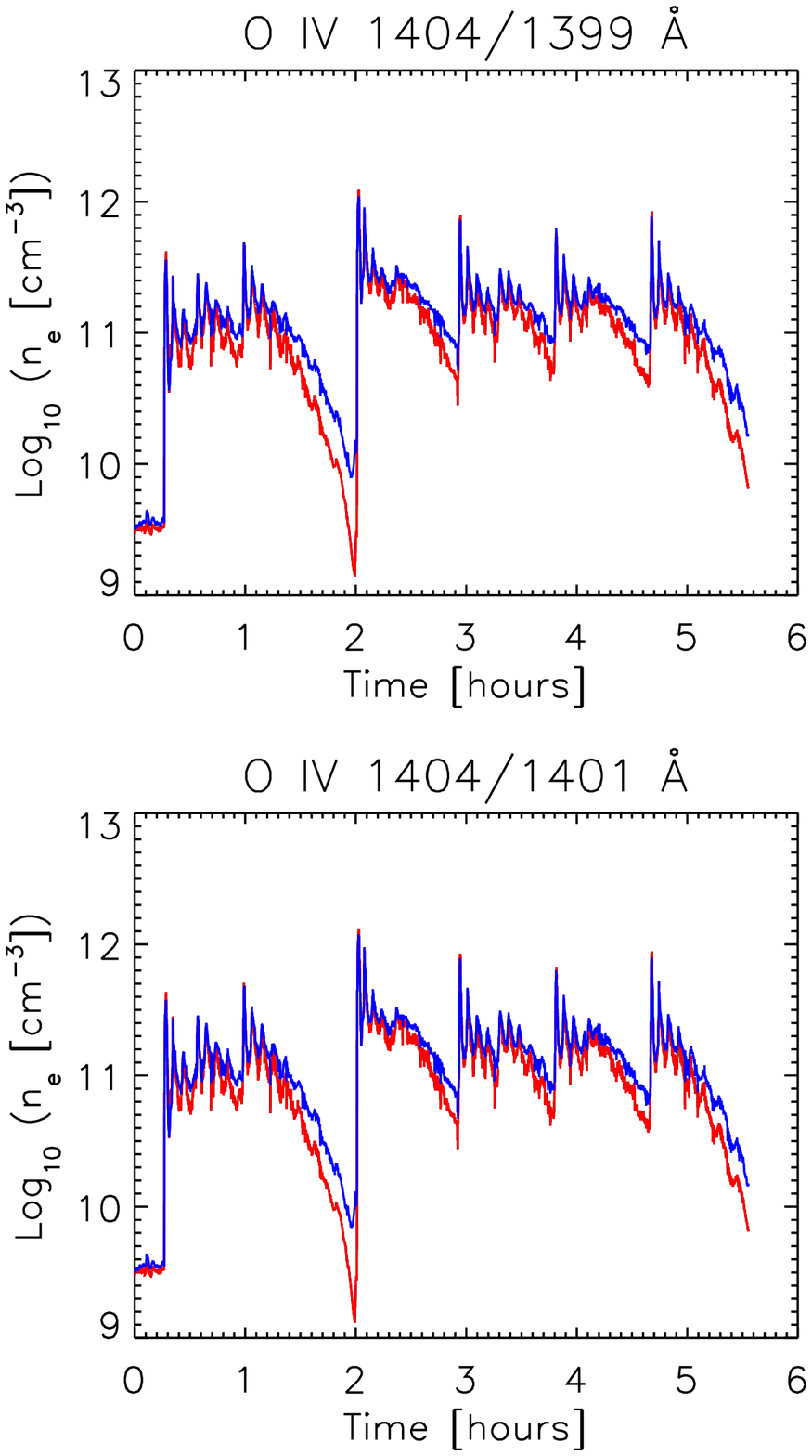}{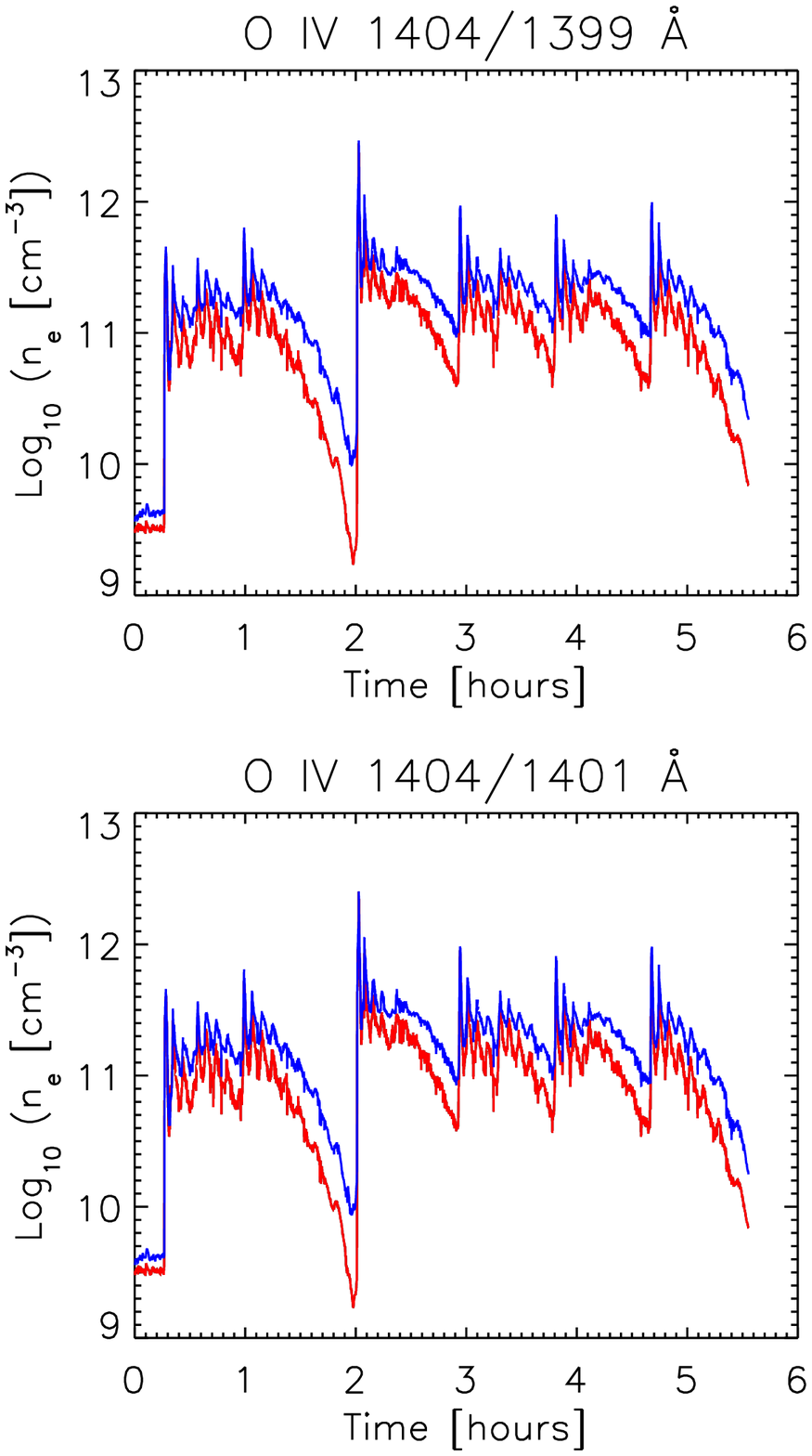}
	\plottwo{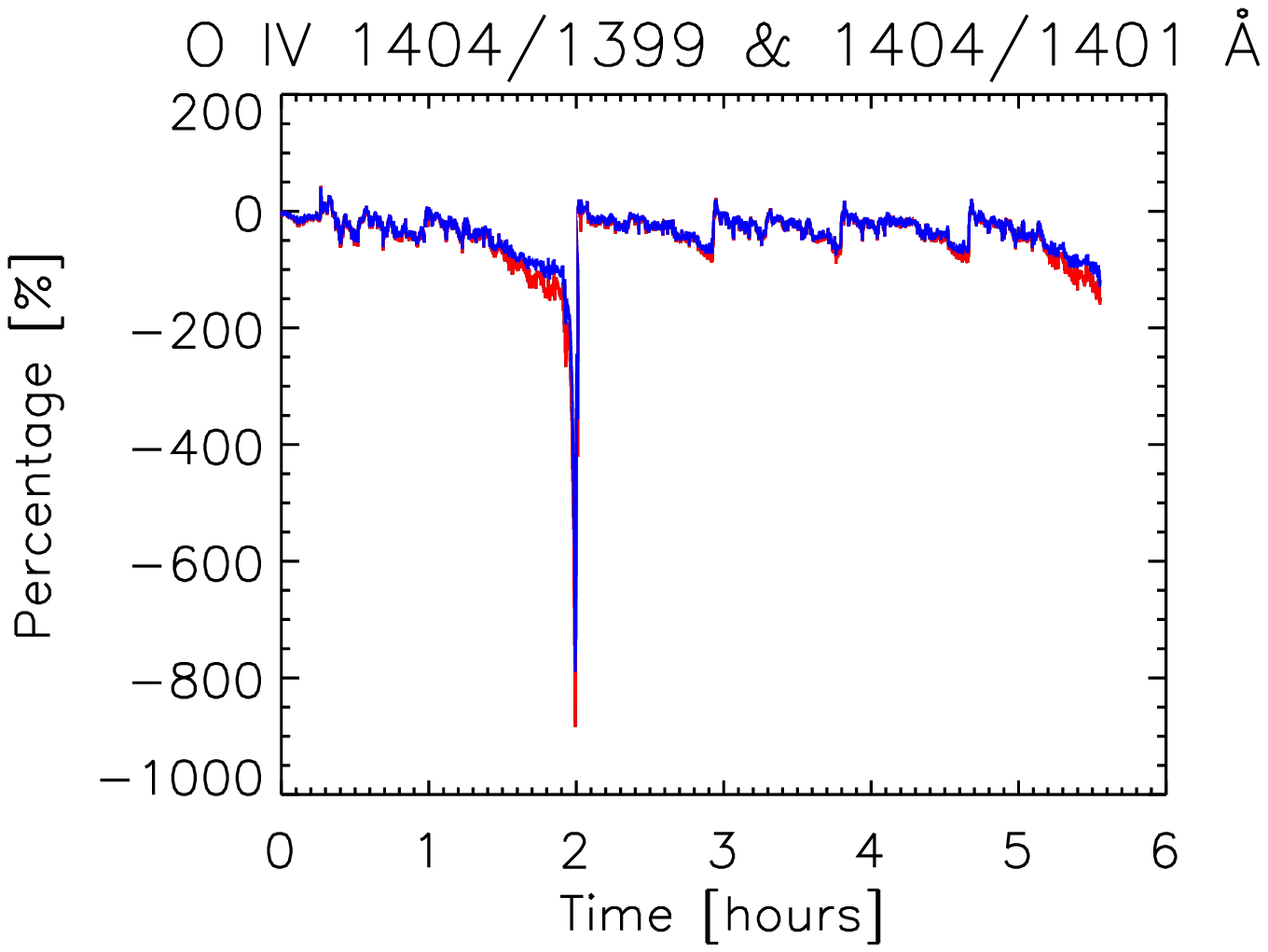}{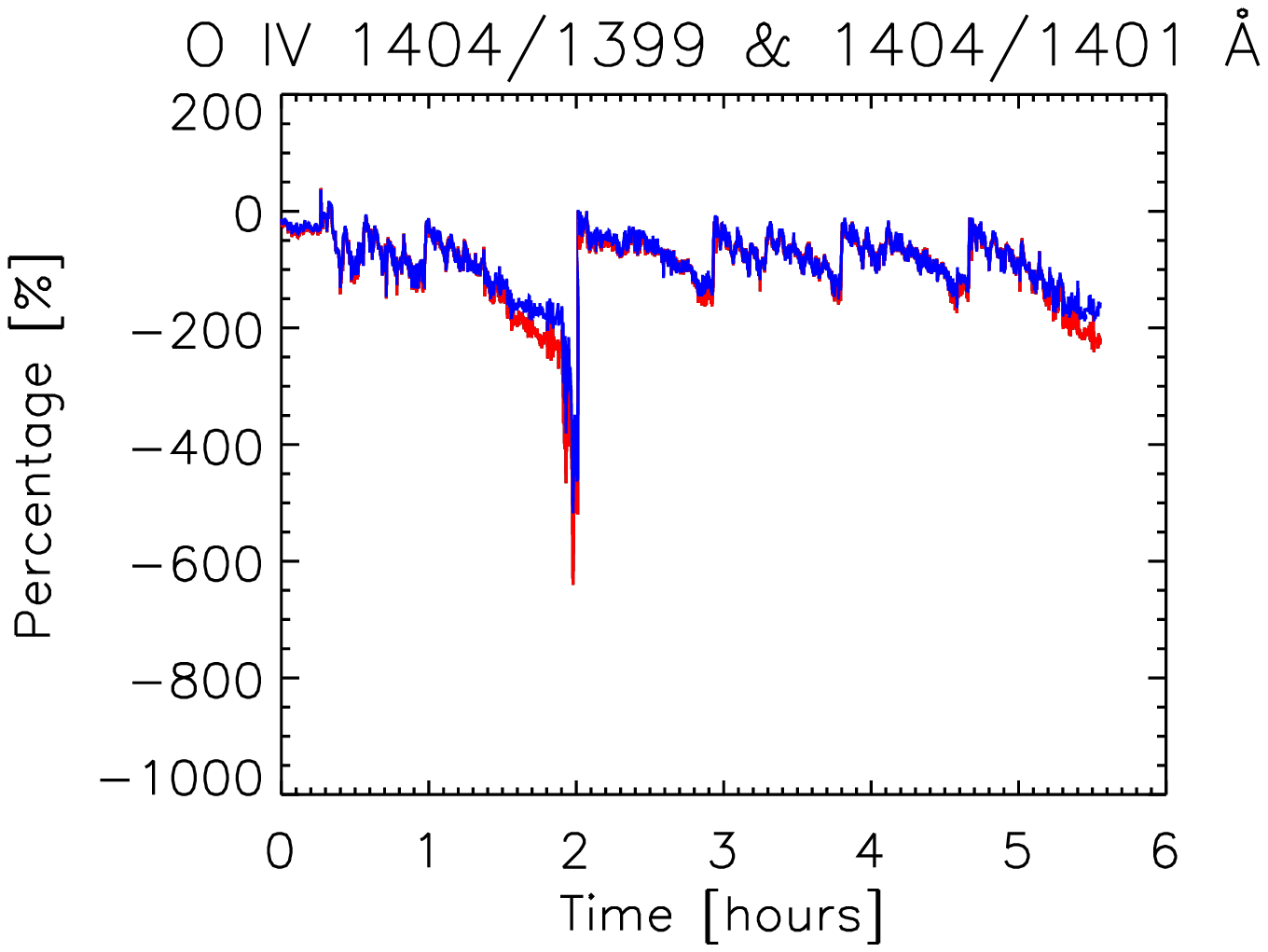}
	\caption{The predicted electron density, calculated from two sets of line-of-sight integrated, density-sensitive, line ratios, in the strong-heating case. The results presented in the left column were calculated in the low-density limit. The results in the right column were calculated when accounting for density dependence. The top (middle) row shows the predicted density with the \oiv\ 1404~/~1399 (1404~/~1401)~\AA~line pair. In each of these plots the red (blue) curve is calculated for an equilibrium (non-equilibrium) ion population. The bottom row shows the difference between each line pair calculated for equilibrium and non-equilibrium ion populations. The red (blue) curve corresponds to the 1399.78 (1401.16)~\AA~pair.}
	\label{fig6}
\end{figure}

Figure~\ref{fig6} shows the predicted electron density calculated from the same pairs of line-of-sight integrated, density-sensitive line ratios used to produce Figure~\ref{fig3}, but now for the strong-heating case. The results broadly follow the weak-heating case: non-equilibrium ionization (blue curves in the upper four panels) produces a higher density than the equilibrium approximation; the density increase is enhanced by the additional effect of density-dependent quenching of dielectronic recombination in the non-equilibrium case; and the 1401.16~\AA~ratio is less susceptible to the combination of non-equilibrium ionization and density dependence (blue curves in the lower two panels). There is a particularly powerful heating event at 2~hours, following a long period of draining, in the strong-heating case. The combined effect of relatively low coronal density $(n_e<10^9~\mbox{cm$^{-3}$})$ and a large volumetric heating rate $(E_H=0.38~\mbox{erg/cm$^3$/s})$ quickly raises the coronal temperature to 15~MK (Figure~\ref{fig1}). Examining Figure~\ref{fig6} one can see that the density predicted in the low-density and density-dependent cases, at the time of this strong heating event, differs by a factor of almost two (in equilibrium and non-equilibrium ionization). A significantly higher density is predicted in the density-dependent case. However, the difference between the densities predicted in equilibrium and non-equilibrium ionization are greatest for the rates in the low-density case. This is best seen in the lower left panel of Figure~\ref{fig6} at 2 hours, where differences of $\simeq800$\% are reached compared to $\simeq600$\% in the density-dependent case. At all other times (and for all of the events in the weak-heating case), it can be seen that the difference between the densities predicted in equilibrium and non-equilibrium ionization are greatest for the density-dependent rates. This isolated case is probably due to the timescale of the local temperature change during the strong heating event becoming shorter than the timescale for transport across the temperature gradient, corresponding to a comparable temperature change, where the ions are carried into higher density plasma.

\begin{figure}
	\plottwo{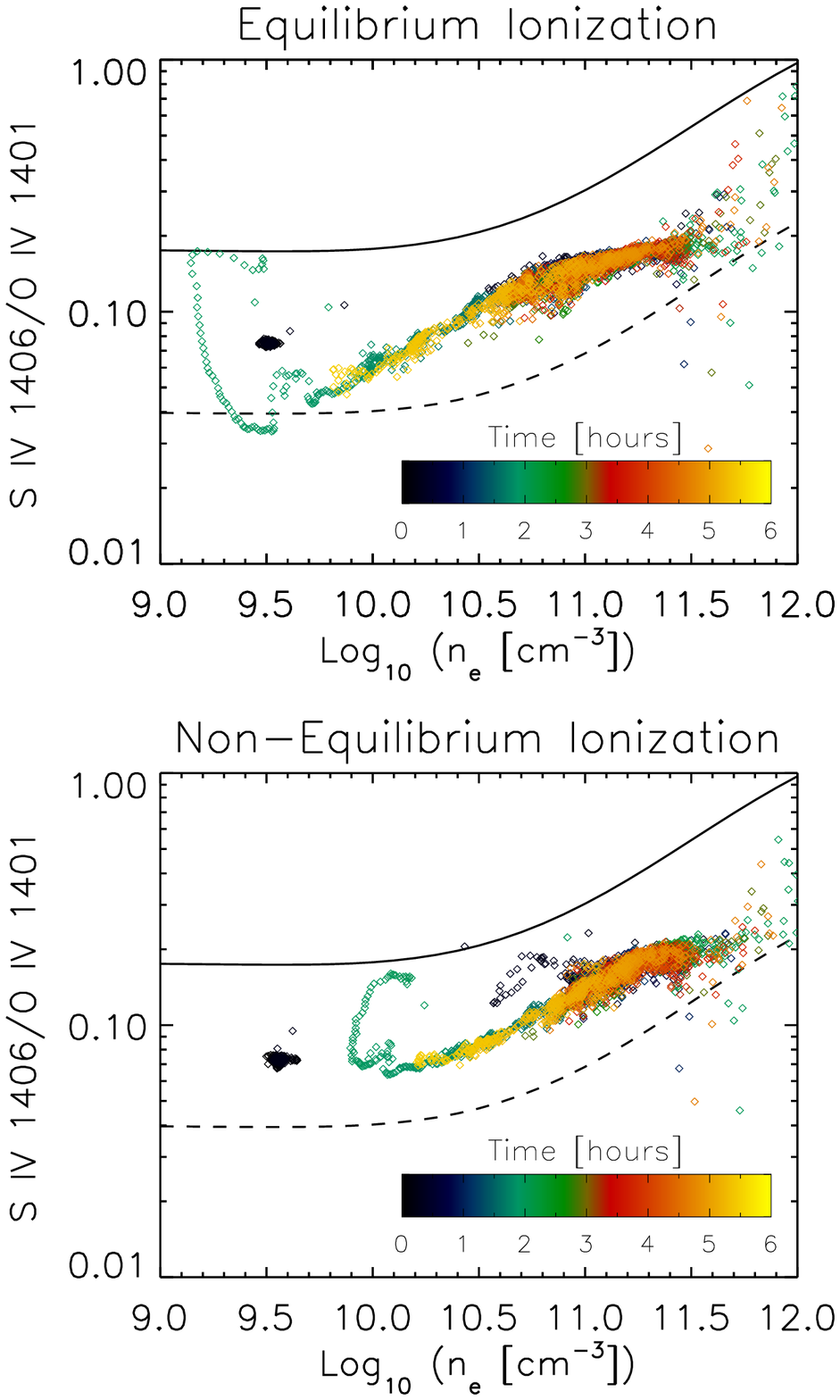}{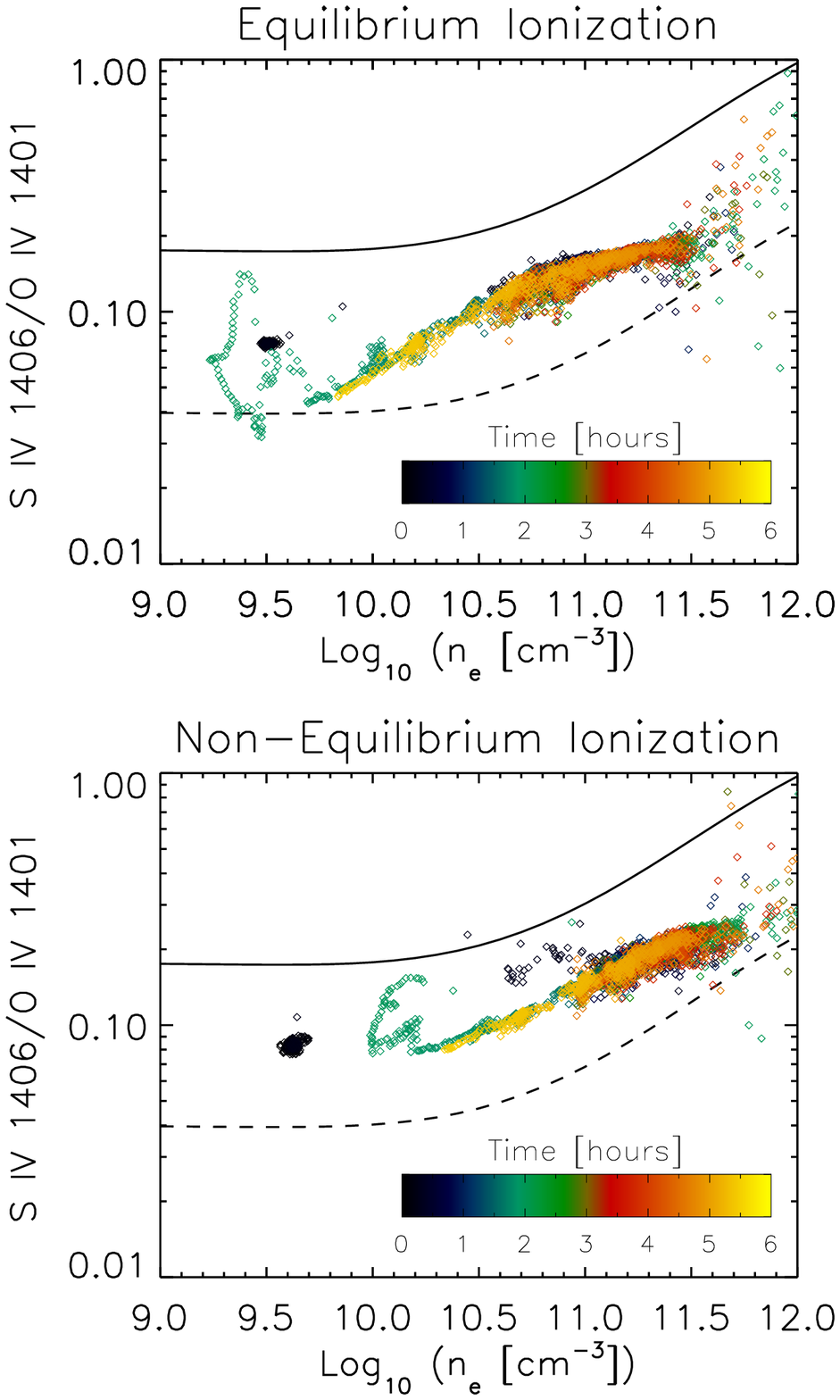}
	\caption{The predicted \siv\ to \oiv\ intensity ratio in the low-density limit (left column) and accounting for density dependence (right column), plotted as a function of electron density, in the strong-heating case. The details of the plot are described in the caption to Figure~\ref{fig4}.}
	\label{fig7}
\end{figure}

Figure~\ref{fig7} shows the equivalent set of plots to Figure~\ref{fig4} for the strong-heating case. Non-equilibrium ionization, again, tends to pull the emitting ions toward higher densities. The effect is most pronounced when comparing the low-density ranges of the corresponding equilibrium and non-equilibrium ionization plots. The spread in the ratio values remains similar across all of the plots. The extended tail at low densities (blueish-green data points) corresponds to the long period of radiative and enthalpy-driven cooling and draining prior to the strong heating event at the 2~hour mark (there are similar colored data points at the highest densities). Intuitively, one might suppose this to be a relatively gentle phase during the evolution of a coronal loop, but the strong down-flows that develop have the same physical consequences for the transition region ionization state as they do during coronal heating, except now they are sustained for a significantly longer period. The emitting ions are transported to regions of substantially higher density and the extended duration of the flows allows a large population to accumulate, such that they dominate emission from the same ions in warmer, overlying regions and line ratio diagnostics report commensurately larger densities. Figure~\ref{fig7} shows differences of up to 0.8~dex in the predicted electron density following a long period of cooling and draining. Density-dependent dielectronic recombination contributes, approximately, an additional 0.1~dex to the density compared to the low-density limit.

\begin{figure}
	\centering
	\includegraphics[scale=0.60]{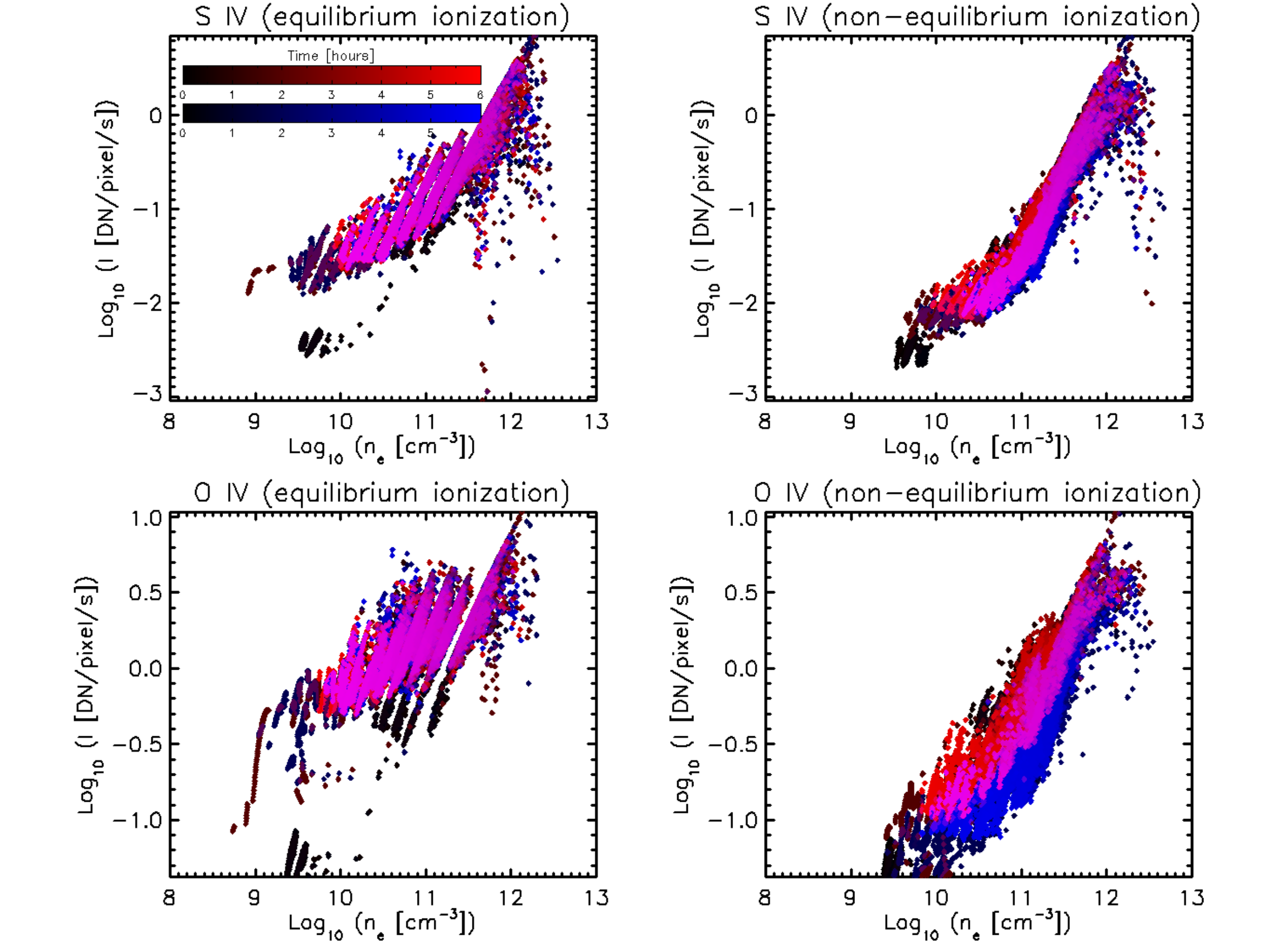}
	\\
	\includegraphics[scale=0.60]{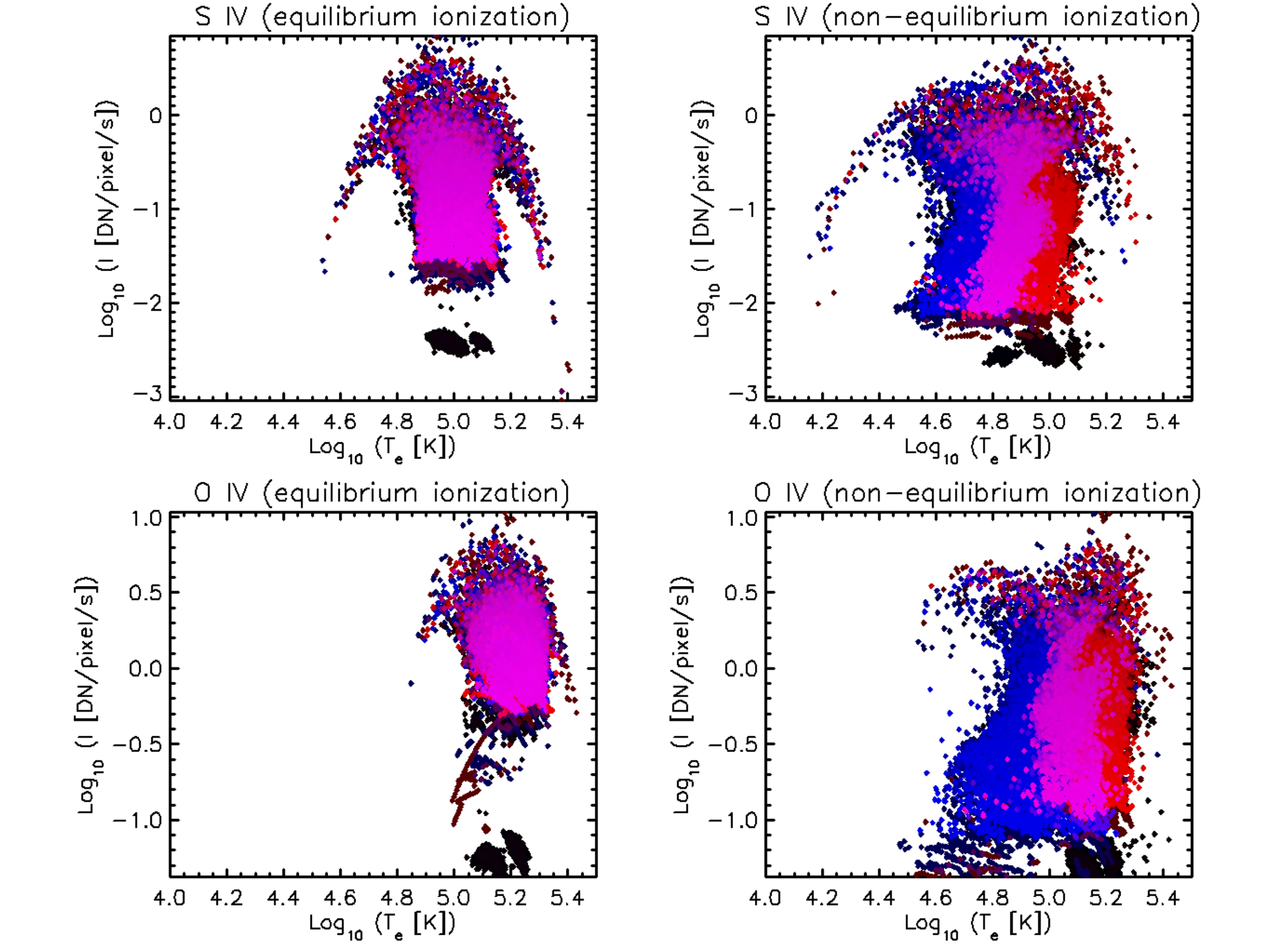}
	\caption{The predicted intensities for the emission lines, plotted as a function of density (upper four panels) and temperature (lower four panels), in the strong-heating case. The details of the plot are described in the caption to Figure~\ref{fig5}.}
	\label{fig8}
\end{figure}

The density and temperature ranges over which the \oiv\ and \siv\ lines emit, equivalent to Figure~\ref{fig5}, for the strong-heating case are shown in Figure~\ref{fig8}. There remains strong overlap between the low-density and density-dependent dielectronic recombination cases in equilibrium ionization (the plots are dominated by purple data points). The trend for decreasing intensity, a shift towards higher densities, and broadening of the temperature distribution towards lower temperatures is also apparent once again in the non-equilibrium ionization case. The intensity decrease is far more pronounced for \oiv\ than for \siv, and the \oiv\ temperature distribution broadens by about 0.6~dex, while the \siv\ distribution broadens by approximately 0.4~dex, compared to the low-density limit. These results are also consistent with the enhanced line-of-sight integrated emission, and associated increased electron densities predicted by line ratio diagnostics, being due to greater emitting column depth, rather than density effects.

\section{Discussion and Conclusions}
\label{discon}

In this paper we have investigated the effects associated with non-equilibrium ionization and density-dependent dielectronic recombination on \iris\ transition region spectral diagnostics. Recent studies have focused on \siiv\ and \oiv\ in non-equilibrium conditions and the effect on the relative intensities of their emission lines \citep{Doyle2013,Olluri2013b,Martinez2016,Dzifcakova2018}. We have focused on the relative intensities of \siv\ and \oiv\ emission lines. In addition to providing a sensitive diagnostic of non-equilibrium ionization, the S/O ratio has the advantage of being less susceptible to chemical abundance variations than the Si/O ratio. Sulphur generally behaves in a manner similar to high-FIP elements such as oxygen \citep[see, for example,][and references therein]{Testa2010}.

We begin the discussion by noting the remarkably good agreement between the predicted \oiv\ intensities in Fig.~\ref{fig2} and the range of observed intensities in Fig.~\ref{fig_obs3} (with no prior-tuning of any parameter values); this is particularly evident at the bright foot-point regions where the measured line intensities reach $\approx30$~DN/pixel/s and only impulsive heating events produce commensurate values in Fig~\ref{fig2}. We emphasize that this investigation does not represent an attempt to reproduce a particular set of observations with a numerical model, but to quantitatively examine the relevant physics (specifically atomic processes) in comparable conditions. The peak \oiv\ intensities predicted by the strong heating experiment are a factor of $\sim3$ greater (O$(100)$~DN/pixel/s) and so early indications are that the weak heating experiment is closer to representing conditions in the four active regions of our study, but this does not preclude significantly different conditions being found in a broader sample of active regions. We know that active regions with a range of peak temperatures exist, for example.

Focusing on our sample of four active regions and further justifying our assertion of weak heating in their particular case, consider Figure~\ref{fig_obs5} which shows the \siv/\oiv\ line ratios plotted as a function of density for each pixel of the observed active regions selected for this study. The data points are color-coded according to the \oiv\ line intensity, where red and white $(>10^4~$DN$)$ indicate the brightest pixels. While there is significant scatter when the full color range is considered, the brightest data points are largely confined to the range $10 \le \log_{10} n_e \le 11.5$. Given that the brightest emission is formed within this density range and the line intensity scales with $n_e^2$, it is reasonable to suppose that it arises as a consequence of increased plasma pressure in the overlying atmosphere (due to heating raising the temperature, for example) compressing the transition region in the magnetic field-aligned direction, thereby increasing the plasma density and the emission line intensity. In fact, the population of white (most bright) data-points are consistently shifted to larger densities relative to the red data-points in Figure~\ref{fig_obs5}. Examining Figures~\ref{fig3} and \ref{fig6} which show the densities predicted by the numerical experiments in the cases of weak- and strong-heating, respectively, it is clear that the weak-heating case most closely recovers the observed density range since the corresponding intensity variations in Figure~\ref{fig2} follow the temporal pattern of the density variations. The brightest emission falls within a density range that overlaps substantially with the observed range. The brightest emission in the strong-heating case would tend to fall predominantly in the range $11 \le \log_{10} n_e \le 12$; about one order of magnitude larger than observed with considerably higher counts. The distribution of predicted densities is broadest for the blue curves shown in the right-hand column of Figure~\ref{fig3} (for both diagnostic ratios), which take account both of density-dependent dielectronic recombination and non-equilibrium ionization. Thus, it seems that both processes (at least) are required to most comprehensively recover the observed range.

Now, consider Figure~\ref{fig_obs5} in the context of Figures~\ref{fig4} (weak-heating) and \ref{fig7} (strong-heating). The data-points follow a line that is roughly intermediate between the upper and lower theoretical ratio curves in the upper-row of plots shown in Figure~\ref{fig4} and in all plots of Figure~\ref{fig7}. The densities are larger in Figure~\ref{fig7} and as a result the data-points clearly remain closer to the equilibrium distribution. The data-points in the lower-row of Figure~\ref{fig4} show a considerably broader distribution in the values of the S/O intensity ratios, which most closely agrees with the distribution of the brightest data-points in Figure~\ref{fig_obs5}. The plot corresponding to S/O ratios calculated for density-dependent dielectronic recombination and non-equilibrium ionization (lower-right plot in Figure~\ref{fig4}) even contains data-points that fall outside the upper theoretical ratio curve, as also seen for some of the brightest data-points in Figure~\ref{fig_obs5}, indicating that non-equilibrium ionization makes the largest contribution to achieving a broad distribution, with a smaller contribution from the density-dependence of the atomic rates. For example, the strongest event in the weak-heating case at 2.7~hours (Figure~\ref{fig1}) corresponds to the green data-points in Figure~\ref{fig4}. Non-equilibrium ionization has scattered them along the S/O intensity ratio axis, relative to the equilibrium ionization case, and density-dependent dielectronic recombination has contributed to lifting a few of them above the upper theoretical ratio curve. The train of similar-sized heating events, between 3 and 4.5 hours, in the weak-heating case correspond to the red and orange data-points in Figure~\ref{fig4}. Again, non-equilibrium ionization has broadened their distribution along the S/O intensity ratio axis, with a small cluster that are slightly enhanced in the density-dependent rate experiment relative to the low-density limit case. As the plasma is heated and allowed to cool down, the \oiv\ emission produces an intensity that varies around $\approx10$~DN/pixel/s in the weak-heating experiment (Figure~\ref{fig2}). Allowing that the total observed intensity is found by summing the measured counts over 16 ($4\times4$) pixels and multiplying by the 60~s exposure time, the total predicted \oiv\ intensity is around $10^4$~DN which is consistent with the brightest data-points in Figure~\ref{fig_obs5}. Consequently, the conclusion is once again that the weak heating case corresponding to density dependence and non-equilibrium ionization provides the best (not necessarily complete) description of the heating properties and key atomic processes in the observed regions.

The discussion so far has focused on the brightest (red and white) cluster of observed pixels shown in Figure~\ref{fig_obs5}, which we have assumed are those in some phase of the heating~/~cooling cycle. However, there is a considerable spread of data-points which our canonical numerical experiments are not able to fully reproduce. The outlying data-points, which contribute most to the large spread, are typically at least an order of magnitude less bright. This raises several possibilities in order to explain their occurrence. The data-points occupying low-density regions of the phase space represented in Figure~\ref{fig_obs5} might simply be due to weaker heating events, where reduced \oiv\ intensities are a natural consequence of the $n_e^2$ scaling. Conversely, strong heating events can act to push data-points into the high-density regions of the space and larger values for the S/O intensity ratio, where the \siv\ intensity is enhanced relative to \oiv\, as shown in Figure~\ref{fig4} (green data-points corresponding to the strong heating event at 2.7~hours in the weak-heating experiment) and in Figure~\ref{fig7} (green data-points corresponding to the strongest event at 2~hours in the strong-heating experiment). The heating and cooling cycle can reproduce the distribution of data-points in the phase space, where the extent of coverage depends on the strength of heating. The increase in coronal pressure during heating drives the transition region downward into higher densities with larger intensity ratios. At the conclusion of heating, the overlying, dense corona rapidly cools and experiences a pressure drop, which leads to a strong rebound of the transition region upward into lower densities where the intensity ratios are reduced. Some of the data-points at higher densities correspond to dimmer pixels, which may be surprising when one considers the $n_e^2$ scaling of the intensity, but indicates that stronger heating could be a plausible explanation if the plasma temperature is increased such that the ion population emits no strong lines in the wavelength range of the instrument.

The data-points that are considerably more difficult to explain with our current modeling effort are those that fall within the range $10 \le \log_{10} n_e \le 11$, but towards and below the lower of the theoretical S/O intensity ratio curves. The data-points in this region require a strong enhancement in the \oiv\ intensity relative to \siv\, which does not readily arise in our numerical experiments and a process by which it might arise does not immediately present itself. One might conjecture that extreme non-equilibrium ionization could potentially broaden the distribution of intensity ratios sufficiently. It is particularly interesting to note that most of the data-points that fall into this regime are among the brightest (green and red), with intensities greater than $10^3$~DN, and so they may very well be associated with coronal heating events. We will investigate this region of the phase space in greater detail and address it in a subsequent article in this series. The dimmest of the population of data-points (black through blue) correspond to pixels with total intensities less than $10^2$~DN (of order 0.1~DN/pixel/s), which indicates that signal to noise issues may be important in the analysis. These data-points tend to be distributed toward higher intensity ratios and at the extremes of the density range, suggesting that low counts are due either to small $n_e^2$ or plasma temperatures close to the limit of instrument sensitivity, and that signal to noise may play a role in the extent of the observed distribution.

We conclude by noting some additional features of the observations which show the models are incomplete in their present form and that we plan to pursue in subsequent investigations. The active regions for which data is plotted in the left-hand column of Figure~\ref{fig_obs5} showed only moderate activity during the observing period and the distribution of the brightest (red) data-points agrees well with the predicted distribution made by the weak-heating experiment when non-equilibrium ionization and density-dependent dielectronic recombination are included in the calculation (Figure~\ref{fig4}, lower-right panel). These data-points are clustered within the density range $10 \le \log_{10} n_e \le 11$ and S/O intensity ratios between 0.1 and 0.2. Significantly greater levels of activity were observed in the active regions shown in the right-hand column of Figure~\ref{fig_obs5} when the observations were made. The brightest data-points (white) are clustered towards higher densities and S/O intensity ratios. While we have established that stronger heating can explain higher densities, Figure~\ref{fig7} shows that it does not necessarily produce larger intensity ratios in the required density range. For example, the ratios at $\log_{10} n_e=11$ remain confined to the range 0.1 to 0.2, whereas in Figure~\ref{fig_obs5} the ratios corresponding to that density in the most active regions can be a factor of 2 larger. Consequently, it cannot only be a matter of the magnitude of the energy input into more active regions that yields larger S/O intensity ratios, other factors related to the activity level and~/~or heating must also be important.

We are left with several questions to explore: (1) Are enhanced S/O intensity ratios consistently observed in active regions of greater activity; (2) Are enhanced S/O ratios connected to properties of the heating such as the frequency (they cannot be due to magnitude of energy input alone); (3) Are thus far unaccounted for processes responsible for the enhanced ratios, such as non-Maxwellian electron velocity distributions due to free-streaming electrons from the hot corona altering the atomic ionization and excitation states; and (4) Are the enhanced ratios due to the FIP-effect mechanism working to greater effect in active regions of greater activity level, thereby increasing the abundance of sulphur (a borderline low-~/~high-FIP element)?

To conclude: The observed properties of the active regions included in our study, particularly the S/O intensity ratio (a powerful diagnostic of non-equilibrium ionization and potentially other atomic processes), are well-explained by a train of relatively weak nanoflares that raise the coronal temperature to $2-3$~MK and its density to $\sim10^9$~cm$^{-3}$ when non-equilibrium ionization and density-dependent dielectronic recombination are accounted for in the emission line synthesis calculations. In the case of equilibrium ionization the density-dependence of the atomic rates does not have a significant effect on the line intensity, but can play a major role at the high-densities that non-equilibrium ionization allows the emitting ions to access.

\acknowledgements
We thank the anonymous referee for a thoughtful review of the manuscript and their helpful comments. Our work also benefited from discussions with Vanessa Polito. SJB and PT were both funded for this work by the NASA Heliophysics Guest Investigator program for the \iris\ mission (grant NNX15AF47G).


\end{document}